\documentclass[preprint]{aastex}
\begin{document}
\title{Multiscale Gaussian Random Fields for Cosmological Simulations}
\author{Edmund Bertschinger}
\affil{Department of Physics, MIT Room 6-207, 77 Massachusetts Ave.,
  Cambridge, MA 02139; \email{edbert@mit.edu}}
\begin{abstract}
This paper describes the generation of initial conditions for numerical
simulations in cosmology with multiple levels of resolution, or multiscale
simulations.  We present the theory of adaptive mesh refinement of Gaussian
random fields followed by the implementation and testing of a computer code
package performing this refinement called {\tt GRAFIC2}.  This package is
available to the computational cosmology community at
{\tt http://arcturus.mit.edu/grafic/} or by email from the author.
\end{abstract}

\section{INTRODUCTION}
\label{sec:intro}

Advances in computational algorithms combined with the steady advance of
computer technology have made it possible to simulate regions of the
universe with unprecedented dynamic range \citep{b98}.  Realistic
simulations of galaxy formation require spatial resolution better than
1 kpc and mass resolution better than $10^6\ M_\odot$ in volumes at least
100 Mpc across containing more than $10^{17}\ M_\odot$.  Cosmologists have
made significant progress towards these requirements.  In simulations of
dark matter halos, \cite{fk97} and \cite{ghigna2000} have achieved more
than 4 orders of magnitude in spatial resolution while the Virgo Consortium
has performed simulations with $10^9$ particles \citep{virgo}.  Recently,
\cite{abn} have performed a simulation of the formation of the first
subgalactic molecular clouds using adaptive mesh refinement with a spatial
dynamic range of 262,144 and a mass dynamic range more than $10^{16}$.

The possibility to resolve numerically such vast dynamic ranges of length
and mass begs the question of what are the appropriate initial conditions
for such simulations.  Hierarchical structure formation models like the
cold dark matter (CDM) family of models have increasing amounts of power at
smaller scales.  This power should be present in the initial conditions.
For simulations of spatially constant resolution, this is straightforward
to achieve using existing community codes \citep{b95}.  However, workers
increasingly are using multiscale methods in which the best resolution is
concentrated in only a small fraction of the simulation volume.  How
should multiscale simulations be initialized?

Many workers currently initialize multiscale models following the approach
of \cite{kqbg94}.  First, a Gaussian random field of density fluctuations
(and the corresponding irrotational velocity field) is sampled on a Cartesian
lattice of fixed spacing $\Delta x$.  Then, $\Delta x$ is decreased by
an integer factor $r>1$ and a new Gaussian random field is sampled with
$r^3$ times as many points, such that the low-frequency Fourier components
(up to the Nyquist frequency $\pi/\Delta x$ in each dimension) agree exactly
with those sampled on the lower-resolution grid.

This method has two drawbacks.  First, it is limited by the size of the
largest Fast Fourier Transform (FFT) that can be performed, since the
Gaussian noise is sampled on a uniform lattice in Fourier space.  This
represents a severe limitation for adaptive mesh refinement codes which
are able to achieve much higher dynamic range.  Second, the uniform
high-frequency sampling on the fine grid is inconsistent with the actual
sampling of the mass used in the evolutionary calculations.  Multiscale
simulations have grid cells, hence particle masses, of more than one size.
The gravitational field produced by a distribution of unequal particle
masses differs from that produced with constant resolution.  In the linear
regime, the velocity and displacement should be proportional to the
gravitational field.  With the method of \cite{kqbg94}, they are not.  We are
challenged to develop a method for sampling multiscale Gaussian random
fields consistent with the multiresolution sampling of mass.

A satisfactory method should satisfy several requirements in addition
to correctly accounting for variable mass resolution.  First, each refined
field should preserve exactly the discretized long-wavelength amplitude
and phase so as to truly refine the lower-resolution sample.  Second,
high-frequency power should be added in such a way that the multiscale
fields are an exact sample from the power spectrum over the whole range
of wavelengths sampled.  Because multiscale fields are not sampled on a
uniform lattice, it is not the power spectrum but rather than spatial
two-point correlation function that should be exactly sampled.  Finally,
a practical method should have a memory requirement and computational
cost independent of refinement so that it is not limited by the size of
the largest FFT that can be performed.

This paper presents the analytic theory and practical implementation
of multiscale Gaussian random field sampling methods that meet these
requirements.  Our algorithms are the equivalent of adaptive mesh
refinement applied to Gaussian random fields.  The mathematical properties
of such fields are simple enough so that an exact algorithm may be developed.
Practical implementation requires certain approximations to be made but
they can be evaluated and the errors controlled.

The essential idea enabling this development is that Gaussian random fields
can be sampled in real space rather than Fourier space (hereafter $k$-space).
Adaptive mesh refinement can then be performed in real space conceptually
just as it is done in the nonlinear evolution code used by \cite{abn}.

How can the long-range correlations of Gaussian random fields be properly
accounted for in real space?  In an elegant paper, \cite{salmon} pointed out
that any Gaussian random field (perhaps subject to regularity conditions
such as having a continuous power spectrum) sampled on a lattice can be
written as the convolution of white noise with a function that we will call
the transfer function.  Salmon recognized the advantages of multiresolution
initial conditions and developed a tree algorithm to perform the convolutions.
Tree algorithms have the advantage that they work for any mesh---regular,
hierarchical, or unstructured.

Next, \cite{pen97} pointed out that FFTs may be used to perform the
convolutions in such a way that the two-point correlations of the
sampled fields are exact, in contrast with the usual $k$-space methods
which produce exact power spectra but not two-point correlations.  The
key is that the transfer functions may be evaluated in real space
accurately at large separation free from distortions caused by the
discretization of $k$-space.  Pen also pointed out that this method
allows the mean density in the box to differ from the cosmic average,
and that the method could be extended to hierarchical grids.

This paper builds upon the work of \cite{salmon} and \cite{pen97}
as well as the author's earlier {\tt COSMICS} package \citep{b95},
which included a module called {\tt GRAFIC} (Gaussian Random Field
Initial Conditions).  {\tt GRAFIC} implemented the standard $k$-space
sampling method for generating Gaussian random fields on periodic
rectangular lattices.  This paper presents the theory and computational
methods for a new package for generating multiscale Gaussian random
fields for cosmological initial conditions called {\tt GRAFIC2}.
This paper contains the fine print for the owner's manual to
{\tt GRAFIC2}, as it were.

This paper is organized as follows.  \S \ref{sec:method} reviews the
mathematical method for generating Gaussian random fields through
convolution of white noise including adaptive mesh refinement.
\S \ref{sec:transf} presents methods for the all-important computation
of transfer functions.  \S \ref{sec:implem} presents important details
of implementation.  Exact sampling requires careful consideration of
both the short-wavelength components added when a field is refined
(\S \ref{sec:short}) as well as the long-wavelength components
interpolated from the lower-resolution grid (\S \ref{sec:long}).
As we show, the long-wavelength components must be convolved with the
appropriate anti-aliasing filter.  Truncation of this filter to a
subvolume (a step required to avoid intractably large convolutions)
introduces errors that we analyze and reduce to the few percent level
in \S \ref{sec:fixv}.

The method is extended to hierarchical grids in \S \ref{sec:multiple}.
\S \ref{sec:tricks} presents additional tricks with Gaussian random
fields made possible by the white noise convolution method.  \S \ref{sec:end}
summarizes results and describes the public distribution of the computer
codes developed herein for multiscale Gaussian random fields.

\section{MATHEMATICAL METHOD}
\label{sec:method}

The starting point is the continuous Fourier representation of the density
fluctuation field:
\begin{equation}
  \label{delft}
  \delta(\vec x\,)=\int d^3k\,e^{i\vec k\cdot\vec x}\,T(k)\xi(\vec k\,)\ ,
\end{equation}
where $\xi(\vec x\,)$ is Gaussian white noise with power spectrum
\begin{equation}
  \label{wiener1}
  \left\langle\xi(\vec k_1)\xi(\vec k_2)\right\rangle=\delta_{\rm D}^3
    (\vec k_1+\vec k_2)\ .
\end{equation}
Here $\delta_{\rm D}(\vec k\,)$ is the Dirac delta function and we are assuming
that space is Euclidean.  The function $T(k)$ is the transfer function
relative to white noise, and it is related simply to the power spectrum of
$\delta(\vec x\,)$:
\begin{equation}
  \label{transferk}
  T(k)=[P(k)]^{1/2}\ .
\end{equation}
Note that $\xi(\vec k\,)$ and $T(k)$ both have units of $[\hbox{length}]^
{3/2}$ and that $T(k)$ is an ordinary function while $\xi(\vec k\,)$ is a
stochastic field (a distribution).

The next step is to recognize that equation (\ref{delft}) can be written
as a convolution \citep{salmon}:
\begin{equation}
  \label{delcon}
  \delta(\vec x\,)=(\xi\ast T)(\vec x\,)=\int d^3x'\,\xi(\vec x^{\,\prime})
    T(\vert\vec x-\vec x^{\,\prime}\vert)
\end{equation}
where
\begin{equation}
  \label{transferx}
  T(\vert\vec x\,\vert)=\int{d^3k\over(2\pi)^3}\,e^{i\vec k\cdot\vec x}\,T(k)
\end{equation}
and
\begin{equation}
  \label{wiener2}
  \left\langle\xi(\vec x_1)\xi(\vec x_2)\right\rangle=(2\pi)^3
    \delta_{\rm D}^3(\vec x_1-\vec x_2)\ .
\end{equation}
The spatial two-point correlation function of $\delta(\vec x\,)$ is simply
$(2\pi)^3(T\ast T)(\vec x\,)$.

Thus, we may construct an arbitrary Gaussian random field by the convolution
of white noise with a convolution kernel determined by the power spectrum.
The white noise process is formally divergent; from equation (\ref{wiener2}),
$\xi(\vec x\,)$ is drawn from a Gaussian distribution with infinite variance.
This strange behavior arises because we are including contributions from
all scales and $\xi(\vec x\,)$ is ultraviolet-divergent.  Physically this
divergence may be cut off by the power spectrum, although the standard cold
dark matter spectrum still leads to a logarithmic divergence of the dark
matter density fluctuations at small scales.  In practice the integral is
cut off at high wavenumber by discretizing space with a finite cell size.

The standard method for generating Gaussian random fields relies on
discretizing equation (\ref{delft}) with a Cartesian mesh in a finite
parallelpiped with periodic boundary conditions.  The spatial dynamic
range is then limited by the size of the largest FFT that can be performed.
The Fourier domain is used because the random variables at different
points are statistically independent aside from the condition $\xi(-\vec
k\,)=\xi^\ast(\vec k\,)$ required to enforce reality of $\delta(\vec x\,)$.
In the spatial domain, $\delta(\vec x\,)$ has long-range correlations
that are difficult to sample unless one first goes to Fourier space.

The velocity field (or displacement field, in the case of dark matter
particles) obeys similar equations; only the transfer function $T(k)$
is modified.

The convolution method described in this paper evaluates the density and
velocity fields using equation (\ref{delcon}) instead of equation
(\ref{delft}).  It relies on the fact that white noise is uncorrelated
in the spatial domain as well as the Fourier domain, hence there is no
difficulty in sampling $\xi(\vec x\,)$.  Once we have such a sample,
it is unnecessary to use a single enormous FFT to evaluate the convolution
equation (\ref{delcon}). Tree algorithms may be used \citep{salmon} or
multiple FFTs with appropriate boundary conditions \citep{pen97}.  The
algorithm we develop extends the ideas of Pen.

\subsection{Discrete Convolution Method Without Refinement}
\label{sec:disconv}

The heart of our method lies in the discretization of equations (\ref
{delft})-(\ref{wiener2}) and their application to density fields
with spatially variable resolution.  The density field is represented
on a hierarchy of nested Cartesian grids so that FFT methods can be
used to perform the convolutions.

Before describing convolution with spatially variable resolution,
we first describe the discrete convolution method for a single grid
of $M$ points per dimension.  For simplicity of presentation we assume
here a cube of length $L$ with periodic boundary conditions, although
the code that implements the convolution is generalized to allow
any parallelpiped.  The grid positions are $\vec x(\vec m\,)
=(L/M)\vec m$ where $\vec m$ is an integer triplet with components
$m_i\in[0,M)$.  Equation (\ref{delft}) becomes
\begin{equation}
  \label{delfft}
  \delta(\vec m\,)=\sum_{\vec\kappa}\exp\left({i2\pi\over M}\vec\kappa\cdot
    \vec m\right)T(k)\xi(\vec k\,)
\end{equation}
where $\vec\kappa=\vec kL/(2\pi)$ is the dimensionless wavenumber; it is
an integer or half-integer triplet with components $\kappa_i\in[-M/2,
M/2)$.  The dimensionless transfer function and spectral noise appearing
in equation (\ref{delfft}) are given by
\begin{equation}
  \label{discpow}
    T(k)\equiv\left[(2\pi/L)^3P(k)\right]^{1/2} ,\ \
    \xi(\vec k\,)=M^{-3}\sum_{\vec m}\exp\left(-{i2\pi\over M}\vec\kappa
    \cdot\vec m\,\right)\,\xi(\vec m\,)\ ,
\end{equation}
where $\xi(\vec m\,)$ is white noise with variance $M^3$:
\begin{equation}
  \label{wiener3}
  \left\langle\xi(\vec m_1)\xi(\vec m_2)\right\rangle=M^3\delta_{\rm K}
    (\vec m_1,\vec m_2)=M^3\times\cases{1\ , &$\vec m_1=\vec m_2$;\cr
      0\ , &$\vec m_1\ne\vec m_2$\ .\cr}
\end{equation}
The subscript K denotes the Kronecker delta.

The discrete convolution algorithm proceeds through the following steps.
\begin{enumerate}
\item Sample $\xi(\vec m\,)$ by generating independent, zero-mean
  normal deviates with variance $M^3$ at each spatial grid point.
\item Use the FFT algorithm to evaluate the second of equations
  (\ref{discpow}).
\item Multiply $\xi(\vec k\,)$ by the discrete transfer function
  $T(k)$.
\item Use the FFT algorithm to evaluate equation (\ref{delfft}).
\end{enumerate}
The result is a discrete approximation to equation (\ref{delft}).

So far, this method is identical to the usual one for generating
Gaussian random fields \citep{b95} except that an extra FFT is
introduced by sampling $\xi(\vec m\,)$ in real space instead of
Fourier space.  This requires more computation but is crucial when
we extend the method to a multiscale hierarchy, which we do next.

\subsection{Mesh Refinement}
\label{sec:meshref}

\unitlength=0.1in
{\samepage
\vspace{0.5in}
\begin{center}
  \begin{picture}(32,32)
  \multiput(0,0)(4,0){9}{\circle*{0.75}}
  \multiput(0,4)(4,0){9}{\circle*{0.75}}
  \multiput(0,8)(4,0){9}{\circle*{0.75}}
  \multiput(0,12)(4,0){9}{\circle*{0.75}}
  \multiput(0,16)(4,0){9}{\circle*{0.75}}
  \multiput(0,20)(4,0){9}{\circle*{0.75}}
  \multiput(0,24)(4,0){9}{\circle*{0.75}}
  \multiput(0,28)(4,0){9}{\circle*{0.75}}
  \multiput(0,32)(4,0){9}{\circle*{0.75}}
  \multiput(9.89,10.07)(1,0){12}{$\times$}
  \multiput(9.89,11.07)(1,0){12}{$\times$}
  \multiput(9.89,12.07)(1,0){12}{$\times$}
  \multiput(9.89,13.07)(1,0){12}{$\times$}
  \multiput(9.89,14.07)(1,0){12}{$\times$}
  \multiput(9.89,15.07)(1,0){12}{$\times$}
  \multiput(9.89,16.07)(1,0){12}{$\times$}
  \multiput(9.89,17.07)(1,0){12}{$\times$}
  \multiput(9.89,18.07)(1,0){12}{$\times$}
  \multiput(9.89,19.07)(1,0){12}{$\times$}
  \multiput(9.89,20.07)(1,0){12}{$\times$}
  \multiput(9.89,21.07)(1,0){12}{$\times$}
  \end{picture}
\end{center}
\begin{figure}[h]
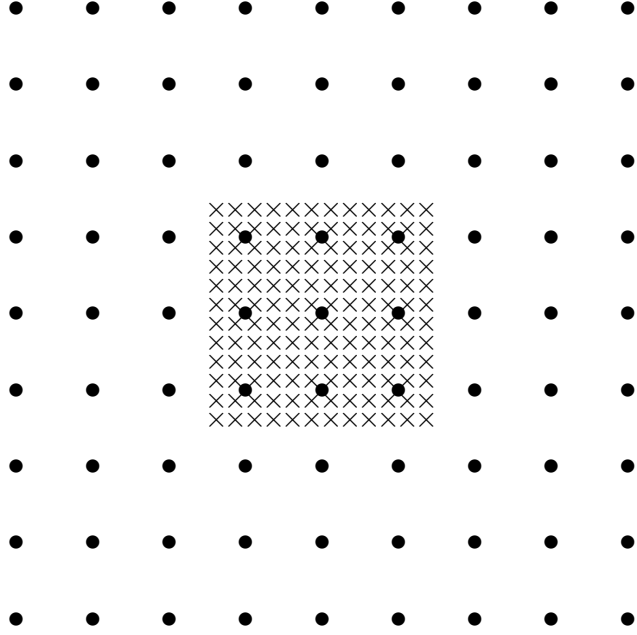

  \caption{An example of mesh refinement with two levels.  The coarse
  grid has size $M=9$, the subvolume has size $M_s=3$, and the
  refinement factor is $r=4$.}
  \label{fig:amr}
\end{figure}
}

Suppose that we have two-level grid hierarchy as shown in Figure
\ref{fig:amr}. Now the spatial grid point positions in the refined
volume are given by two integer triplets, $\vec m$ for the coarse
grid and $\vec n$ for the subgrid:
\begin{equation}
  \label{xgrid}
  \vec x(\vec m,\vec n\,)=\vec x_o+\left(L\over M\right)\left(\vec m+
    {1\over r}\vec n\,\right)\ .
\end{equation}
The subgrid is refined by an integer factor $r>1$, with
$n_i\in[0,r)$.  An offset $\vec x_o=-(r-1)L/(2rM)(1,1,1)$ is applied to
center the refinement.  As a result of mesh refinement, each coarse
grid cell is split up into $r^3$ subcells.

Suppose that we already have a sample of white noise on the coarse
grid, $\xi_0(\vec m\,)$.  Convolution by the appropriate transfer
function using equations (\ref{delfft}) and (\ref{discpow}) then
gives the density field $\delta(\vec m\,)$.  To refine the sampling,
we generate a mesh-refined white-noise sample $\xi(\vec m,\vec n\,)$ and
convolve it with a higher-resolution transfer function.

The refined white-noise sample $\xi(\vec m,\vec n\,)$ should retain
the same low-frequency structure as the coarse-grid sample $\xi_0(\vec m
\,)$.  We ensure this by choosing $\xi(\vec m,\vec n\,)$ to be a sample
of Gaussian white noise subject to the linear constraint
\begin{equation}
  \label{xicon}
  \sum_{\vec n}\xi(\vec m,\vec n\,)=r^3\xi_0(\vec m\,)\ .
\end{equation}
The constraint is easy to apply using the Hoffman-Ribak algorithm
\citep{hr91}. One simply generates an unconstrained white
noise sample $\xi_1(\vec m,\vec n\,)$ with variance $(rM)^3$ and then
applies a linear correction to enforce equation (\ref{xicon}):
\begin{equation}
  \label{xisamp}
  \xi(\vec m,\vec n\,)=\xi_1(\vec m,\vec n\,)+\xi_0(\vec m\,)-
    \bar\xi_1(\vec m\,)\ \ \hbox{where}\ \ \bar\xi_1(\vec m\,)\equiv
    r^{-3}\sum_{\vec n}\xi_1(\vec m,\vec n\,)\ .
\end{equation}
The sample so generated is Gaussian white noise satisfying the constraint
equation (\ref{xicon}) and having the desired covariance
\begin{equation}
  \label{xicov}
  \left\langle\xi(\vec m_1,\vec n_1)\xi(\vec m_2,\vec n_2)
  \right\rangle=(rM)^3\delta_{\rm K}(\vec m_1,\vec m_2)\delta_{\rm K}
  (\vec n_1,\vec n_2)\ .
\end{equation}

Equation (\ref{xisamp}) has a simple interpretation.  Mesh refinement
takes place by splitting each coarse cell (labelled by $\vec m\,$)
into $r^3$ subcells.  The coarse-grid white noise value $\xi_0$ is
first spread to each of the subcells, then a high-frequency correction
$\xi_1-\bar\xi_1$ is added.

\subsection{Subgrid Convolution}
\label{sec:subgrid}

Our method requires performing several convolutions over the subgrid.
In this subsection we describe the method for a generic high-resolution
convolution, which is first expressed in Fourier space as follows:
\begin{equation}
  \label{congen1}
  \delta(\vec m,\vec n\,)=\sum_{\vec k}\exp[i\vec k\cdot\vec x(\vec m,
    \vec n\,)]\,T(k)\xi(\vec k\,)\ .
\end{equation}
The sum is taken over the extended Fourier space of size ($rM)^3$.
This Fourier space extends to wavenumbers $r$ times greater than that
of equation (\ref{delfft}).  We can write the wavevector using two
integer (or half-integer) triplets, $\vec\kappa$ and $\vec b$:
\begin{equation}
  \label{kgrid}
  \vec k=\left(2\pi\over L\right)\left(\vec\kappa+M\vec b\,\right)
\end{equation}
where $\kappa_i\in[-M/2,M/2)$ and $b_i\in[-(r-1)/2,(r-1)/2]$.  The
set of all $\vec\kappa$ for a given $\vec b$ is called a Brillouin
zone.  The coarse grid corresponds to the fundamental Brillouin
zone, $\vec b=(0,0,0)$.  Mesh refinement extends the coverage of
wavenumber space by increasing the number of Brillouin zones to $r^3$
where $r$ is the refinement factor.

The major technical challenge of our algorithm is to perform the
convolution of equation (\ref{congen1}) without storing or summing
over the entire Fourier space.  This is possible when $\delta(\vec m,
\vec n\,)$ is required over only a subgrid in the spatial domain.
The first step is to note that equation (\ref{congen1}) is equivalent to
\begin{equation}
  \label{congen2}
  \delta(\vec m,\vec n\,)=\sum_{\vec m',\vec n^{\,\prime}}\xi
    (\vec m^{\,\prime},\vec n^{\,\prime})T(\vec m-\vec m^{\,\prime},
    \vec n-\vec n^{\,\prime})
\end{equation}
where
\begin{equation}
  \label{congen3}
  \xi(\vec m,\vec n\,)=\sum_{\vec k}\exp[i\vec k\cdot\vec x(\vec m,\vec n\,)]
    \,\xi(\vec k\,)\ ,\quad
  T(\vec m,\vec n\,)=(rM)^{-3}\sum_{\vec k}\exp[i\vec k\cdot\vec x(\vec m,
    \vec n\,)]\,T(k)\ .
\end{equation}

Now, mesh refinement is performed only over the subgrid of size
$(rM_{\rm s})^3$ where $M_{\rm s}<M$, so it is not necessary to evaluate
$\xi(\vec m,\vec n\,)$ and $T(\vec m,\vec n\,)$ for all $(rM)^3$
high-resolution grid points.  We set $\xi(\vec m,\vec n\,)=0$ outside of
the subgrid volume.  Consequently, $T(\vec m,\vec n\,)$ needs to be
evaluated only to distances of $\pm rM_{\rm s}$ grid points in each dimension
in order that all contributions to $\delta(\vec m,\vec n\,)$ be included.
We will describe how the transfer functions $T(\vec m,\vec n\,)$ are
computed in the next subsection.

The function $\xi(\vec m,\vec n\,)$ must also be evaluated on the subgrid.
Because this is a sample of white noise, we simply draw independent
real Gaussian random numbers with zero mean and variance $(rM)^3$ at each
grid point.  Subtracting a mean over coarse grid cells (eq. \ref{xisamp})
or imposing any other desired linear constraints (using the Hoffman-Ribak
method) is easily accomplished.

Once we have $\xi(\vec m,\vec n\,)$ and $T(\vec m,\vec n\,)$ on the
subgrid, the next step is to Fourier transform them.  For simplicity
in presentation, let us suppose that the subgrid is cubic with $N_{\rm
s}=(2rM_{\rm s})^3$ where $M_s$ is the number of coarse grid points
that are refined in each dimension.  The result is
\begin{equation}
  \label{congen4}
  \xi'(\vec k^{\,\prime})={1\over N_{\rm s}}\sum_{\vec x}\exp-(i\vec k^
    {\,\prime}\cdot\vec x\,)\,\xi(\vec x\,)\ ,\quad
  T'(\vec k^{\,\prime\,})=\sum_{\vec x}\exp(-i\vec k^{\,\prime}\cdot
    \vec x\,)\,T(\vec x\,)
\end{equation}
where the sums are taken over the fine grid points in the subvolume,
which has been doubled to the accommodate periodic boundary conditions
required by Fourier convolution.  Primes are placed on the wavevectors
and on the transformed quantities to distinguish them from the original
quantities $\xi(\vec k\,)$ and $T(k)$.  Note that the sampling of
$k$-space is different in equations (\ref{congen3}) and (\ref{congen4})
because the length of the spatial grid has changed from $L$ to $(2M_s/M)L$.
Also, $T'(\vec k^{\,\prime\,})$ is in general not spherically symmetric
even if $T(\vec k\,)$ is spherical.

The final step is to perform the convolution by multiplication in the
subgrid $k$-space followed by Fourier transformation back to real space:
\begin{equation}
  \label{congen5}
  \delta(\vec x\,)=\sum_{\vec k^{\,\prime}}\exp(i\vec k^{\,\prime}\cdot
    \vec x)\,T'(\vec k^{\,\prime\,})\xi(\vec k^{\,\prime})\ .
\end{equation}
The reader may verify that equations (\ref{congen4}) and (\ref{congen5})
give results identical to equations (\ref{congen1}) and (\ref{congen2}),
when $\xi(\vec x\,)$ is zero outside of the subvolume.  Thus, we have
achieved the equivalent of convolution on a grid of size $(rM)^3$ by
using a (typically smaller) grid of size $(2rM_s)^3$.  Note well
that $T'$ is not the same as $T$, because it is based on spatially
truncating $T(\vec m,\vec n\,)$ and making it periodic on a grid of
size $(2M_s/M)L$ instead of $L$.

So far the method looks straightforward.  However, some practical
complications arise which will discuss later, in the computation
of the transfer functions in real space (\S \ref{sec:transf}) and
in the split of our random fields into long- and short-wavelength
parts on the coarse and fine grids, respectively (\S \ref{sec:long}).

Finally, we note that we will perform the convolution of equation
(\ref{congen1})  using a grid of size $2rM_s$ in each
dimension in the standard way using FFTs without requiring
periodic boundary conditions for the subgrid of size $rM_s$.
We calculate the transfer functions in the first octant of size
$(rM_s)^3$ and then reflect them periodically to the other octants using
reflection symmetry (odd along the direction of the displacement,
otherwise even).  In order to achieve isolated boundary conditions,
the white noise field is filled in one octant and set to zero
in the other octants.  If we desire to have periodic boundary conditions
(e.g. for testing), we can set $M_s=M/2$ and fill the full refinement
grid of size $(2rM_s)^3$ with white noise.

\subsection{Computation of Transfer Functions}
\label{sec:transf}

The convolution method requires calculating the transfer functions
$T(\vec x\,)$ for density, velocity, etc., on a high resolution
grid $\vec x(\vec m,\vec n\,)$ of extent $\pm rM_s$ grid points in
each dimension.  The transfer function is given in the continuous
case by equations (\ref{transferk}) and (\ref{transferx}) and in
the discrete case by the first of equations (\ref{discpow}) and the
second of equations (\ref{congen3}).  Our challenge is to compute
the transfer functions on the subgrid without performing an FFT of
size $(rM)^3$, under the assumption that the problem is too large
to fit in the available computer memory.  Also, we wish to avoid
a naive summation of the second of equations (\ref{congen3}), which
would require $O(r^6M_s^3M^3)$ operations.  In practice, $r$ will
be a modest-size integer (from 2 to 8, say) while $M_s$ will be
much larger, of order $M/r$.

We present three solutions to this challenge.  The first two are based,
respectively, on three-dimensional discrete Fourier transforms while
the third is based on a spherical transform.

\subsubsection{Exact Method}
\label{sec:exact}

The first method is equivalent to the second of equations (\ref{congen3})
and is therefore exact in the sense of yielding the same transfer functions
as if we had used full resolution on a grid of size $(rM)^3$.  Note that
\cite{pen97} would call this method approximate because, after the FFT
to the spatial domain, the results differ from the exact spatial transfer
function of equation (\ref{transferx}).  We will say more about this in
\S \ref{sec:spherical}, but note simply that the discretization of $k$-space
required for the FFT makes it impossible for the transfer function to be
exact in both real space and $k$-space.  The transfer function of this
subsection is exact in $k$-space and is equivalent to the usual $k$-space
sampling method.

We rewrite the second of equations (\ref{congen3}) as
\begin{equation}
  \label{transkx1}
  T(\vec m,\vec n\,)=\sum_{\vec \kappa}e^{i(2\pi/M)\vec\kappa\cdot\vec m}
    \,T(\vec\kappa,\vec n\,)\ ,
\end{equation}
where
\begin{equation}
  \label{transkx2}
  T(\vec\kappa,\vec n\,)=(rM)^{-3}\sum_{\vec b}\exp\left[i\left(
    2\pi\over rM\right)(\vec\kappa+M\vec b\,)\cdot\vec n\,\right]
    \,T(\vec k\,)\ .
\end{equation}
The Fourier space is split into Brillouin zones according to
equation (\ref{kgrid}).  Beware that the symbol $T$ has three
different uses here which are distinguished by its arguments: it is
either the transfer function in real space $T(\vec m,\vec n\,)$,
the transfer function in Fourier space $T(\vec k\,)$, or else the
mixed Fourier/real case $T(\vec\kappa,\vec n\,)$.

Equation (\ref{transkx1}) is a simple FFT of size $M^3$.  This is the same
size as is used for generating the coarse grid initial conditions, so it is
tractable.  However, we save the results only at those coarse grid points
$\vec m$ that lie in the refinement subvolume, discarding the rest.  By
performing some unnecessary computation, the FFT reduces the number of
operations required to compute this sum for all $\vec n$ from
$O(M_s^3 M^3)$ to $O(r^3M^3\log M)$, a substantial savings.

Equation (\ref{transkx2}) is also a FFT, in this case of size
$r^3$.  However, we cannot evaluate both equations (\ref{transkx1})
and (\ref{transkx2}) using FFTs without storing $T(\vec\kappa,
\vec n\,)$ for all $(rM)^3$ points.  In order to reduce the storage
to a tractable amount (no more than the larger of $M^3$ and $8r^3M_s^3$),
we must perform an outer loop over $\vec n$ to evaluate $T(\vec m,
\vec n\,)$.  For each $\vec n$, we must compute $T(\vec\kappa,
\vec n\,)$ for all $\vec\kappa$, requiring direct summation in
equation (\ref{transkx2}).  The operations count for all $\vec n$
is then $O(r^6M^3)$, which dominates over the $O(r^3M^3\log M)$
for equation (\ref{transkx1}).  The operations count for equation
(\ref{transkx2}) can be reduced by a factor of up to 6 by using
symmetries when $T(\vec k\,)$ is spherically or azimuthally symmetric.
Nonetheless, if we use this method, computation of the transfer
functions is generally the most costly part of the whole method.

\subsubsection{Minimal $k$-space Sampling Method}
\label{sec:minimal}

If the transfer function falls off rapidly with distance in real space,
there is another way to evaluate $T(\vec m,\vec n\,)$ that is much faster.
It is based on noting that the Fourier sum is an approximation to the
Fourier integral, and another approximation is given by simply changing the
discretization in $k$-space.  In equation (\ref{congen3}), the step size
in $k$-space is $2\pi/L$ where $L$ is the full size of the simulation
volume.  If we increase this step size to $(M/2M_s)2\pi/L$, the transfer
function $T(\vec k\,)$ will be evaluated with exactly the sampling needed
for $T'(\vec k^{\,\prime\,})$ in equation (\ref{congen5}).  In this case
we don't even need to transform $T(\vec k\,)$ to the spatial domain,
truncate and periodize on the subgrid to give $T(\vec m,\vec n\,)$, and
then transform back to get $T'(\vec k^{\,\prime\,})$.  We simply replace
$T'$ with $T$ in Fourier space.  This is exactly equivalent to decreasing
the $k$-space resolution in equation (\ref{congen3}) to the minimum needed
to sample $T(\vec m,\vec n\,)$ on the subgrid.

This method is extremely fast but its speed comes with a cost.  Low
wavenumbers are sampled poorly compared with equations (\ref{transkx1}) and
(\ref{transkx2}), and the transfer functions are truncated in a cube of
size $(2M_s/M)L$ instead of $L$.  The decreased $k$-space sampling leads
to significant real-space errors for distances comparable to the size
of the box.  This may be tolerable for the density but is unacceptable
for the velocity transfer function.  In \S \ref{sec:long} we will introduce
anti-aliasing filters for the coarse grid for which the minimal $k$-space
sampling method is well-suited.  In \S \ref{sec:multiple} we will revisit
the use of the minimally sampled transfer function for the density field.

\subsubsection{Spherical Transform Method}
\label{sec:spherical}

Another fast method can be used when $T(\vec k\,)$ is spherically
symmetric, as it is for the density and the radial component of
displacement.  In this case we approximate the second of equations
(\ref{congen3}) as a continuous Fourier integral,
\begin{eqnarray}
  \label{transkx3}
  T(\vec x\,)&\approx&(rM)^{-3}\int{d^3k\over(\Delta k)^3}\,
    e^{i\vec k\cdot\vec x}\,T(k)\nonumber \\
  &=&\left(L\over2\pi rM\right)^3\int {\sin kr\over kr}
    \,T(k)\,4\pi k^2dk\nonumber \\
  &=&\left(L\over2\pi rM\right)^3{2\pi\over r}\,\hbox{{\cal I}m}
    \!\int_{-\infty}^\infty kT(k)e^{ikr}\,dr\ .
\end{eqnarray}
Aside from units, this is essentially the same as equation
(\ref{transferx}).  The last integral in equation (\ref{transkx3})
can be performed by truncating the Fourier integral at the Nyquist
frequency of the subgrid, $k_{\rm Ny}=\pi rM/L$ and then using a
one-dimensional FFT.  The method is much faster than equations
(\ref{transkx1}) and (\ref{transkx2}).

The  Fourier integral of equation (\ref{transkx3}) can be evaluated
accurately, yielding an essentially exact transfer function in real
space.  This approach was advocated by \cite{pen97}.  However, this is
not necessarily the best approach for cosmological simulations with
periodic boundary conditions.  In order to achieve periodic boundary
conditions, such simulations compute the gravitational fields with
$k$-space discretized at low frequencies as in the second of equations
(\ref{congen3}).  In this case it would be inconsistent to use equation
(\ref{transkx3}) on the top-level grid with periodic boundary
conditions---the displacement field would not be proportional, in the
linear regime, to the gravity field computed by the Poisson solver of
the evolution code.  However, the spherical method is satisfactory for
refinements without periodic boundary conditions.

\subsubsection{Anisotropic Transfer Functions}
\label{sec:aniso}

In order to use equation (\ref{transkx3}), the transfer function
must be spherically symmetric.  This seems natural for the
density field given that the power spectrum $P(k)$ is isotropic.
However, the standard FFT-based method for computing samples
of the density field violates spherical symmetry through the
Cartesian discretization of $k$-space.  As we noted above,
periodic boundary conditions are inconsistent with spherical
symmetry on the largest scales.  Moreover, the displacement transfer
function is multiplied by a factor $i\vec k/k^2$ which breaks
spherical symmetry for each Cartesian component.

\begin{figure}[t]
  \begin{center}
    \includegraphics[scale=0.4]{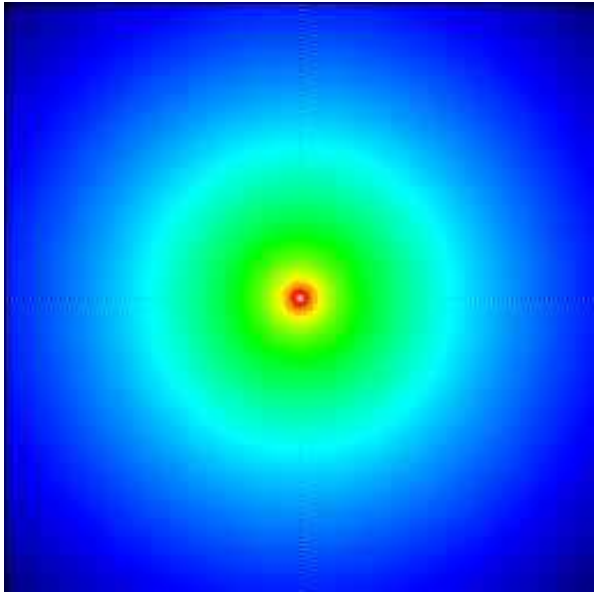}
  \end{center}
  \caption{A slice through the center of the density transfer function in
  real space (``square root of the correlation function'').  The exact
  method of \S \ref{sec:exact} has been used with an unfiltered power
  spectrum.  The cosmological model is flat $\Lambda$CDM and the box is
  64 Mpc across.  False colors are scaled to the logarithm of the transfer
  function, which shows 6 orders of magnitude.  Anisotropy of the
  discrete Fourier transform leads to anisotropic features that are
  barely visible along horizontal and vertical axes through the center.}
  \label{fig:transd3}
\end{figure}

To examine the first concern, namely the non-isotropic discretization
in Fourier space, we examine the transfer function computed
using the exact method of \S \ref{sec:exact}.
Figure \ref{fig:transd3} shows the result for the flat $\Lambda$CDM
model ($\Omega_{\Lambda}=0.65$, $h=0.65$, $\sigma_8=1.0$) with a
$r=4$ refinement of a $M_{\rm s}=32$ subgrid of a $M=256$ grid.
The coarse grid spacing is 1 Mpc.  In effect, the transfer function
has been computed at $1024^3$ resolution on a grid of spacing 0.25 Mpc,
but is shown only within a central region 64 Mpc across ($2M_{\rm s}
\times\hbox{1 Mpc}$).

There is a slight banding visible along the $x$- and $y$-axes in
Figure \ref{fig:transd3}.  The amplitude of this banding ranges from
a relative size of about 20\% at small $r$ to more than a factor of two
at the edges (where the transfer function is very small); however, it
is much smaller away from the coordinate axes.  This anisotropic structure
arises because, although $T(k)$ is spherically symmetric, the Fourier
integration is not carried out over all $\vec k$ but rather only within
a cube of size $2\pi rM/L$.  The Fourier space is periodic (because the
real space is discrete), which breaks the spherical symmetry of $T(\vec x\,)$.
In this case it is the anisotropy at large $k$ that produces the anisotropy
in real space.

This anisotropy is present in the initial conditions generated with the
with the COSMICS package \citep{b95}.  The author's rational for allowing
it was that it is preferable to retain all the power present in the
initial density fluctuation field.  Including all power in the Fourier
cube gives the best possible resolution at small scales while producing
a modest anisotropy along the coordinate axes.  However, the effects of
the anisotropy are unclear and should be more carefully evaluated.
In \S \ref{sec:short}, we will show how the density transfer function
can be made isotropic by filtering.   To determine whether the anisotropy
of unfiltered initial conditions causes any significant errors, full
nonlinear numerical simulations should be performed with and without
filtering.  That test is beyond the scope of this paper.

Additional considerations arise when calculating the transfer function
for the linear velocity or displacement fields.  (The linear velocity and
displacement are proportional to each other.)  The displacement field
$\vec\psi(\vec x\,)$ is related to the density fluctuation field by
$\vec\nabla\cdot\vec\psi=-\delta(\vec x\,)$ in real space or
$\vec T_\psi(\vec k\,)=(i\vec k/k^2)T(k)$ in $k$-space.  Each component
of the displacement field is anisotropic.  This presents no difficulty
for the discrete methods of \S\S \ref{sec:exact}--\ref{sec:minimal}.
There is one subtlety of implementation, however: in Fourier space,
the displacement field must vanish on the Brillouin zone boundaries.
That is, the component of $\vec T_\psi$ along $\vec e_x$ must vanish
on the surfaces $k_x=\pm k_{\rm Ny}$ and similarly for the other
components.  This is required because each component of $\vec T_\psi$ is
both odd and periodic.

If the density transfer function is filtered so as to be
spherical in real space, then the displacement field is radial in real
space and we can obtain the radial component simply by applying Gauss's law:
\begin{equation}
  \label{gauss}
  T_\psi(r)=-{1\over r^2}\int_0^r T(r')r'^2dr'\ .
\end{equation}
The radial integral can be performed from a tabulation of the spherical
density transfer function in real space, $T(r)$, by integrating a cubic
spline or other interpolating function.  The Cartesian components of
displacement follow simply from $\vec T_\psi=T_\psi(r)\vec e_r$.

In summary, we will use the spherical method in the case of spherical
transfer functions, otherwise we will use one of the discrete methods.
If the transfer function is sufficiently localized in real space so that
the Fourier space may be coarsely sampled, the minimal $k$-space sampling
method may be used.  In all cases we will compare against the exact method
to test the accuracy of our approximations.

\section{IMPLEMENTATION}
\label{sec:implem}

In this section we present our implementation of the two-level
adaptive mesh refinement method described in \S \ref{sec:method}
and we discuss the split of our fields into long- and short-wavelength
parts on the coarse and fine grids, respectively.

The high-resolution density field is the superposition of two parts:
\begin{equation}
  \label{del12}
  \delta(\vec m,\vec n\,)=\tilde\delta_0(\vec m,\vec n\,)+\delta_1
    (\vec m,\vec n\,)
\end{equation}
where
\begin{equation}
  \label{del12a}
  \tilde\delta_0=\xi_0(\vec m\,)*T\ ,\quad
  \delta_1=\left[\xi_1(\vec m,\vec n\,)-\bar\xi_1(\vec m\,)\right]*T\ .
\end{equation}
The convolution operator $*$ is defined by equation (\ref{congen2}),
with the transfer function $T(\vec m,\vec n\,)$ defined on a high-resolution
grid.  The net density field is the superposition arising from the
coarse-grid white noise sample $\xi_0(\vec m\,)$ and its high-frequency
correction $\xi_1(\vec m,\vec n\,)-\bar\xi_1(\vec m\,)$ as in equation
(\ref{xisamp}).  Basically, we split the density field (and similarly the
displacement and velocity fields) into long-wavelength and short-wavelength
parts.  In this section we first describe the computation of the
short-wavelength part $\delta_1$, followed by the long-wavelength part
$\tilde\delta_0$.

\subsection{Short Wavelength Components}
\label{sec:short}

\begin{figure}[t]
  \includegraphics[scale=0.4]{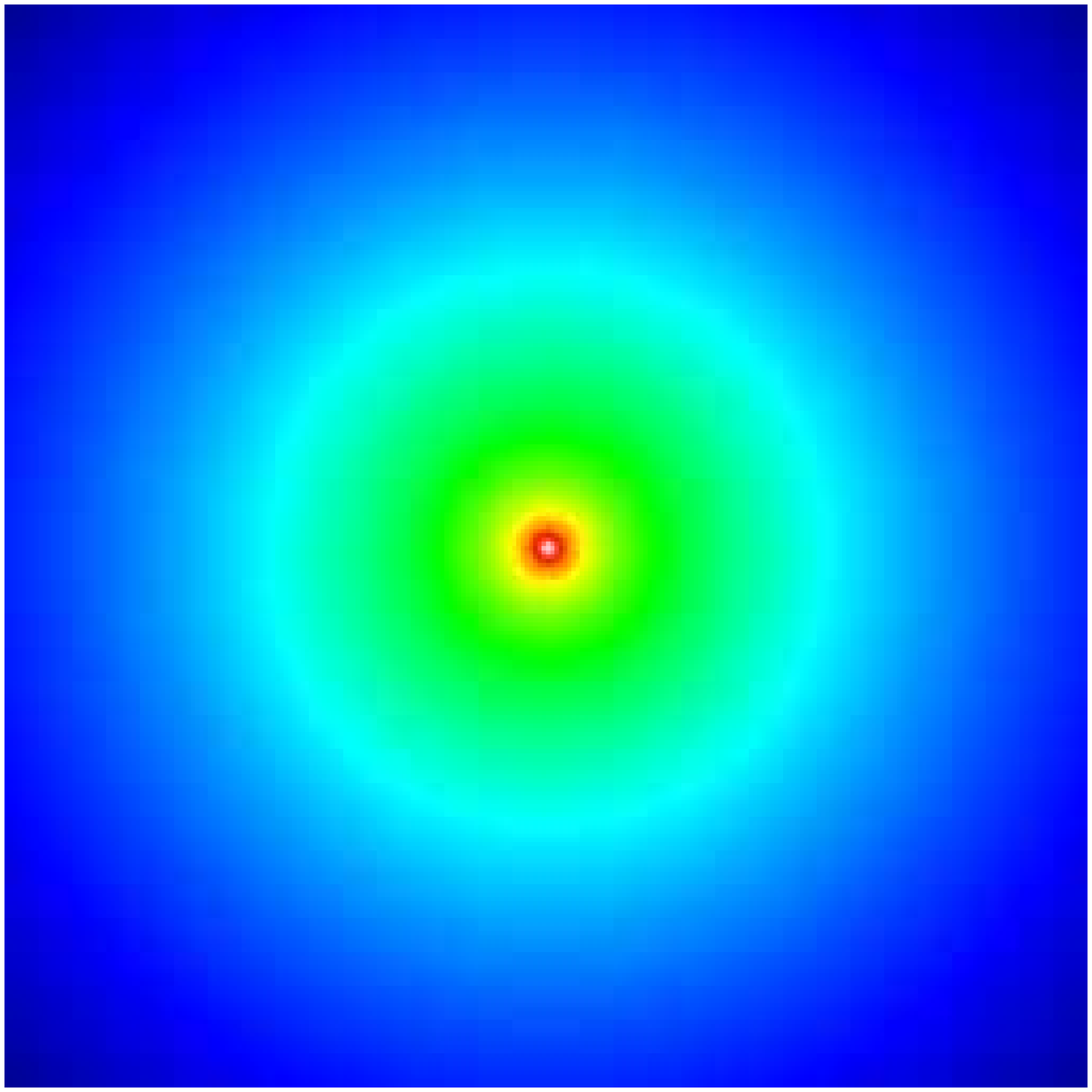}
  \includegraphics[scale=0.4]{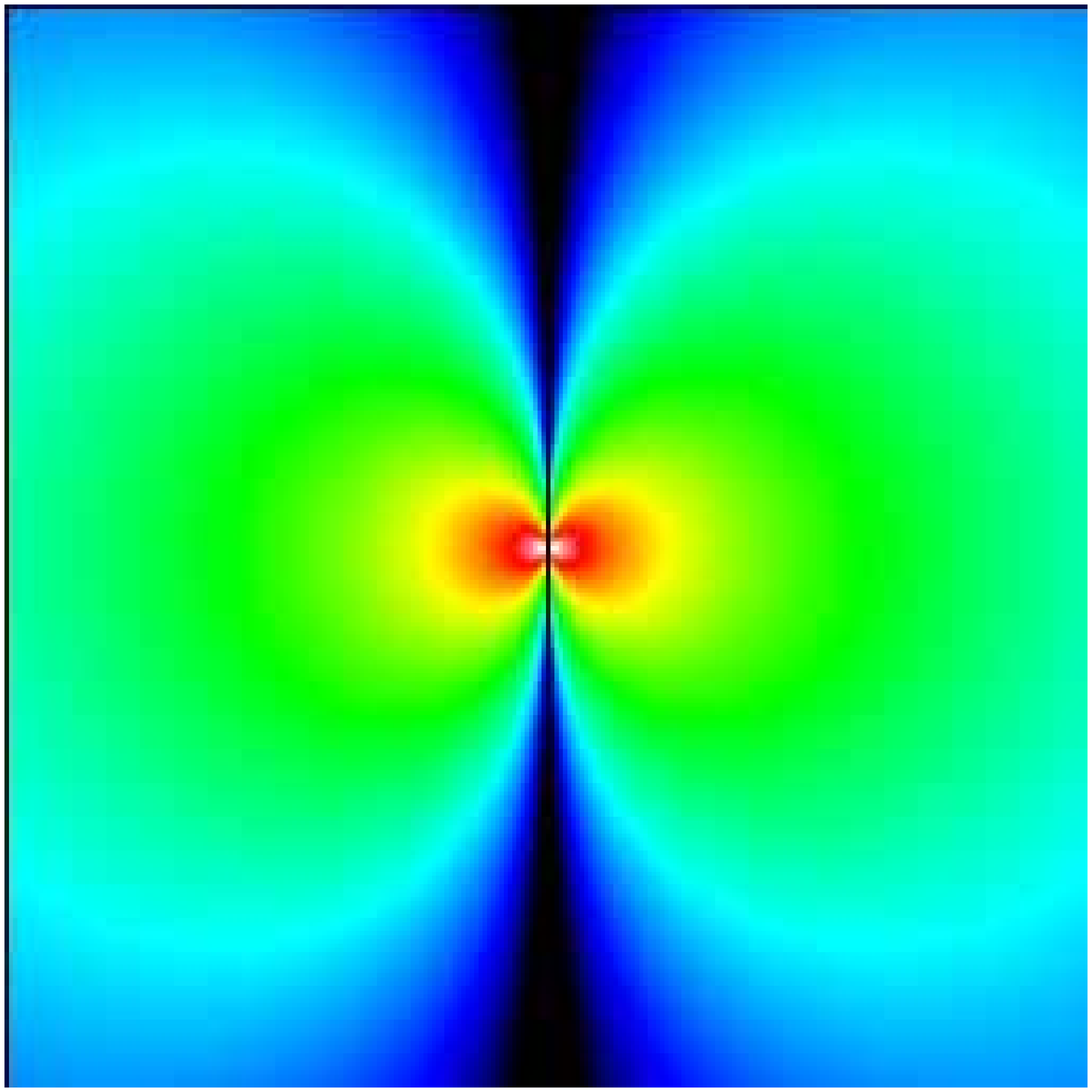}
  \caption{A slice through the center of the transfer functions
  in real space for the density field (left) and one component of the
  displacement field (right).  The model parameters are the same as
  in Figure \ref{fig:transd3} except that a spherical Hanning filter
  (cosine in Fourier space) has been applied to reduce the anisotropy
  that was seen in Figure \ref{fig:transd3}.
  False colors show the logarithm of the transfer function, with 6 orders
  of magnitude shown for the density and 3 orders of magnitude for the
  displacement.  (The absolute value of the displacement is shown; it is
  negative in the right half of the image.)  When convolved with white
  noise, these transfer functions give the density fluctuation and
  $x$-displacement fields in linear theory at redshift $z=0$.}
  \label{fig:transf}
\end{figure}

The high-frequency part of the density field, $\delta_1=(\xi_1-\bar
\xi_1)\ast T$, is straightforward to calculate using the methods of
\S\S \ref{sec:subgrid} and \ref{sec:transf}.  Let us first
consider the transfer functions for the density and displacement
fields, which we show in Figure \ref{fig:transf}.  In order to
eliminate the anisotropy appearing in Figure \ref{fig:transd3},
we have applied a spherical Hanning filter, multiplying $T(k)$ by
$\cos(\pi k/2k_{\rm Ny})$ for $k<k_{\rm Ny}$ and zeroing it
for $k>k_{\rm Ny}$ where $k_{\rm Ny}=\pi rM/L$.  This filter has
removed the anisotropic structure and has also smoothed the density
field near $\vec x=0$.  We have used the spherical method of \S
\ref{sec:spherical}.  The exact method gives results that are visually
almost indistinguishable, with maximum differences of order one percent
because the Hanning filter does completely eliminate the anisotropy of
the discrete Fourier transform.

Each Cartesian component of the displacement transfer function
displays a characteristic dipole pattern because of the projection
from radial motion: $\psi_x=(x/r)\psi(r)$.  A density enhancement
at the origin is accompanied by radial infall (with $\psi_x$ changing
sign across the origin).  Note that the displacement transfer function
falls off much less rapidly with distance than the density transfer
function, illustrating the well-known fact that the linear velocity
field has much more large-scale coherence than the density field.
The linear displacement, velocity, and gravity fields are all proportional
to each other, so one may also interpret $\vec T_\psi$ as the transfer
function for the gravitational field.

\begin{figure}[t]
\includegraphics[scale=0.4]{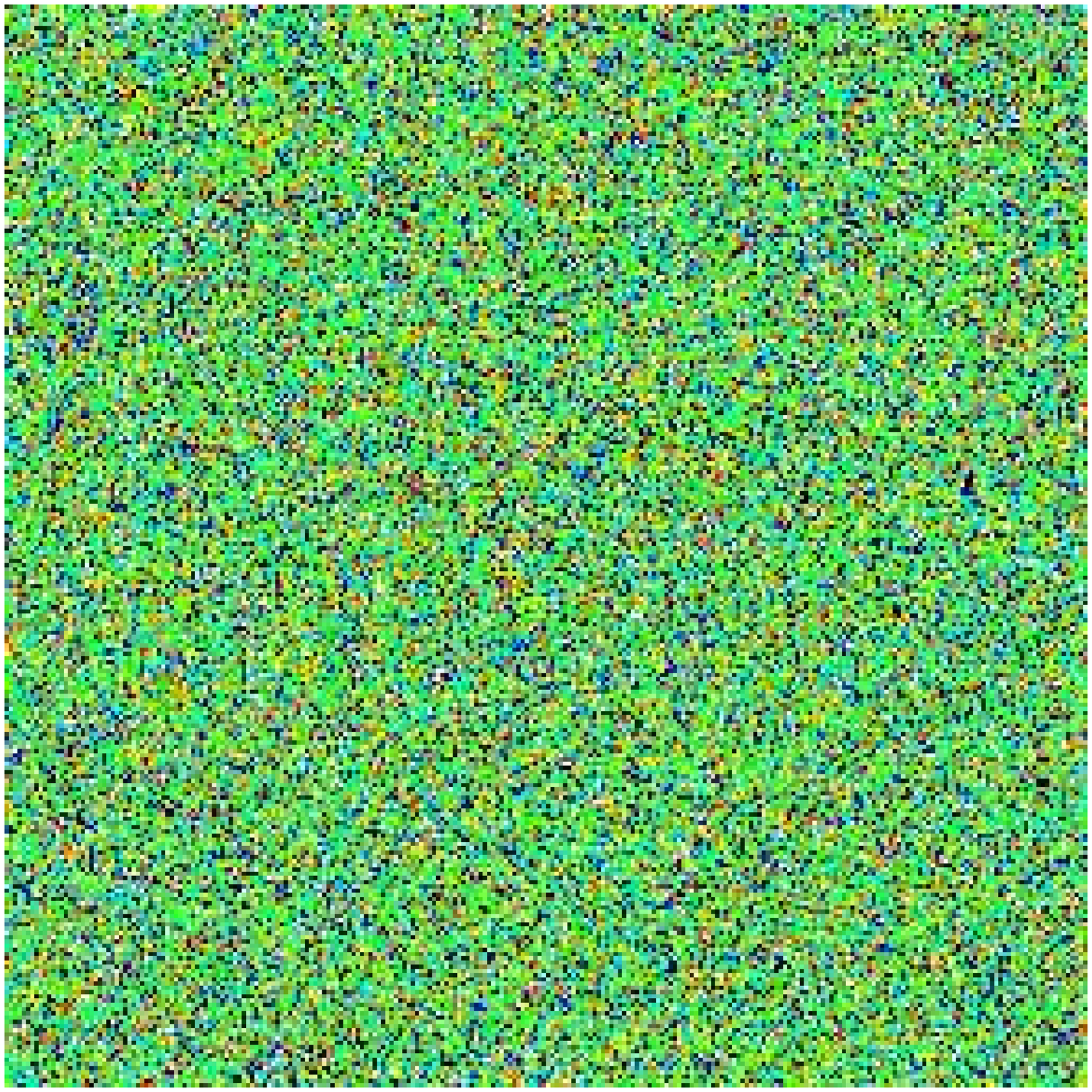}  
\includegraphics[scale=0.4]{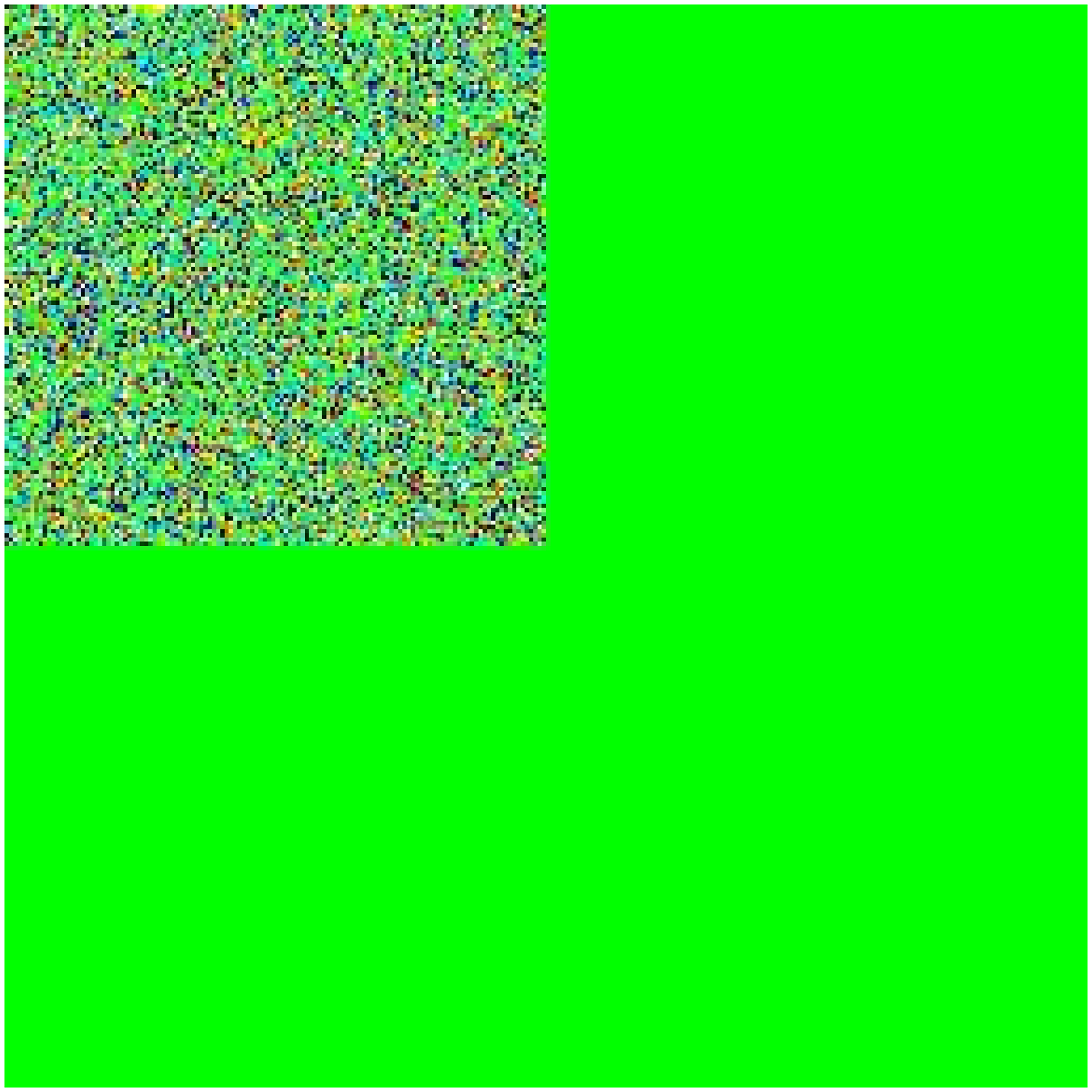}  
  \caption{Slices through a sample of Gaussian noise in a cube of size 64 Mpc.
  Left: pure white noise $\xi_1(\vec m,\vec n\,)$.  Right: zero-padding is
  used for isolated boundary conditions, and the means have been subtracted
  over cells of size 1 Mpc so that we show $\xi_1(\vec m,\vec n\,)-\bar\xi_1
  (\vec m\,)$.  False colors are scaled to linear values ranging from $\pm2$
  standard deviations of $\xi_1$.  The zero level is light green.  Because
  of the predominance of high-frequency power, the filtering applied to the
  right-hand image is not apparent, but the two samples differ in the upper
  left quadrant.}
  \label{fig:wnsamp}
\end{figure}

The next step in computing the subgrid contribution to the initial
conditions is to generate an appropriate sample of Gaussian noise.
Figure \ref{fig:wnsamp} shows two samples of white noise for the
high-resolution subgrid.  The left sample is pure white noise $\xi_1
(\vec m,\vec n\,)$, which is the correct noise sample if we wish to
generate a Gaussian random field with periodic boundary conditions on a
grid of size 64 Mpc (the full width that is shown).  The right sample is
nonzero only in a region 32 Mpc across, as is appropriate for isolated
boundary conditions in a subgrid, and the means over coarse grid cells
have been subtracted, i.e. we plot $\xi_1(\vec m,\vec n\,)-\bar\xi_1
(\vec m\,)$.  This is the appropriate noise sample for computing the
short-wavelength density field $\delta_1$.

\begin{figure}[t]
\includegraphics[scale=0.4]{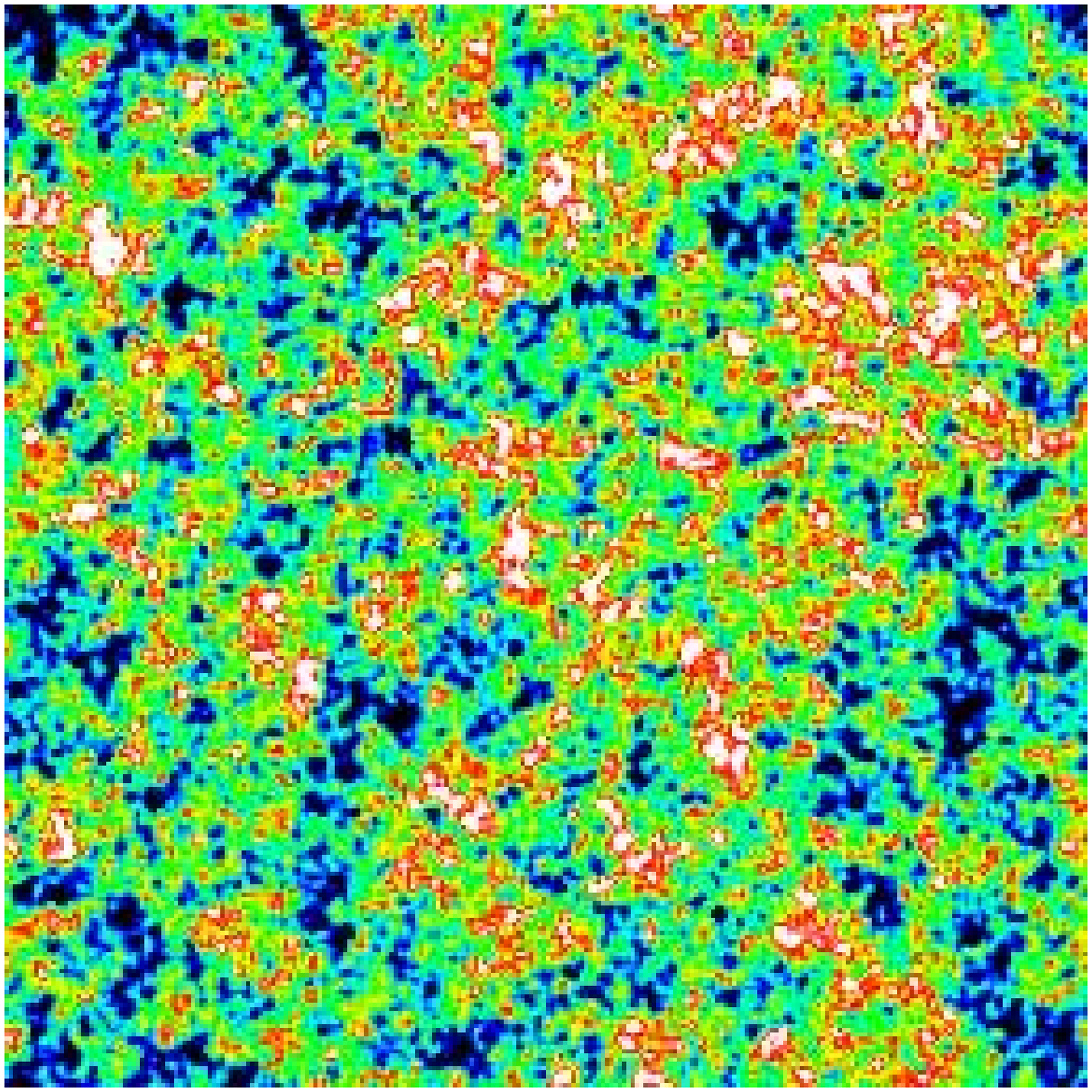}  
\includegraphics[scale=0.4]{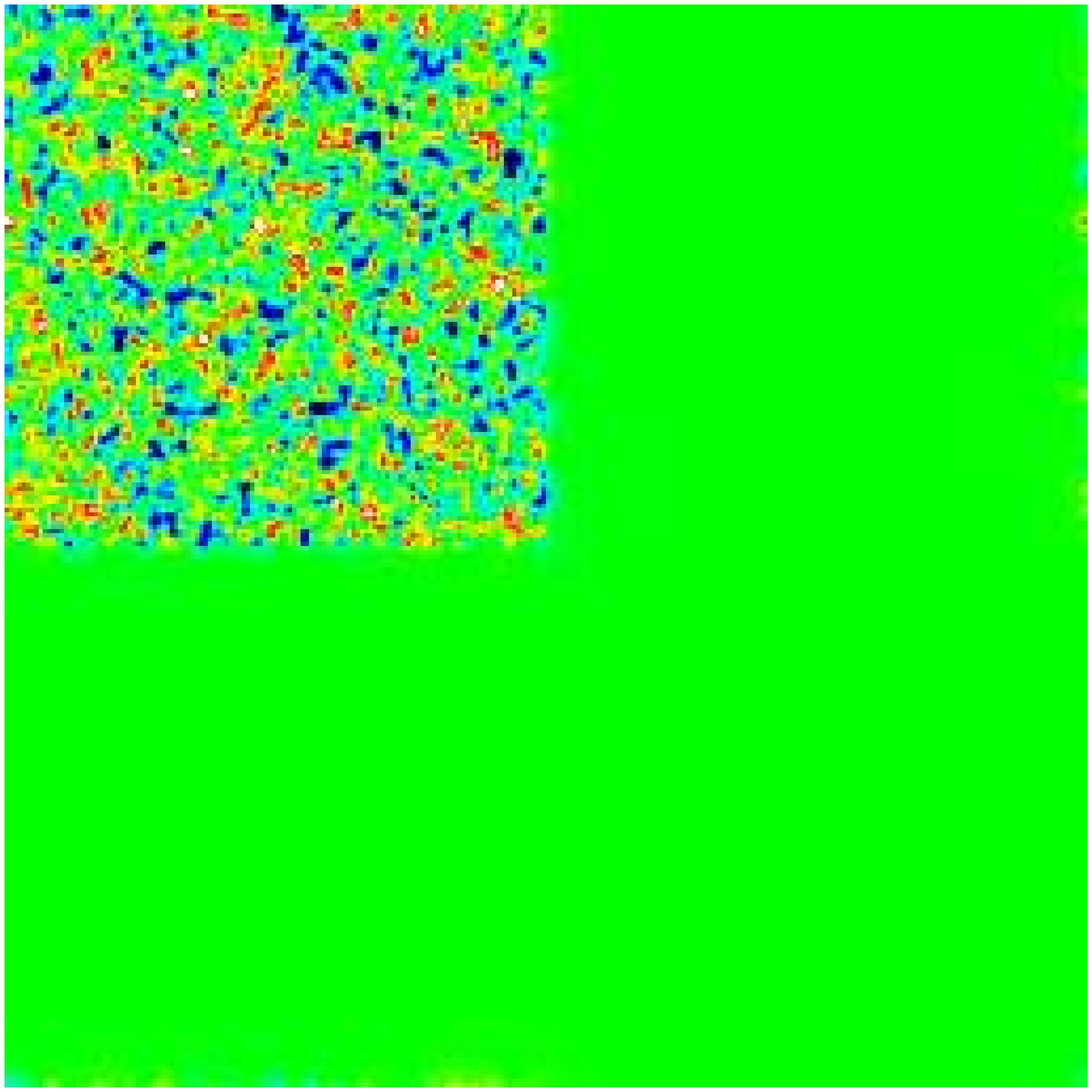}  
  \caption{High resolution density fluctuation field obtained by
  convolving white noise (Figure \ref{fig:wnsamp}) with the $\Lambda$CDM
  density transfer function (left panel of Figure \ref{fig:transf}).  False
  colors are scaled to linear values ranging from $\pm2$ standard deviations
  for the left panel.  The left and right panels correspond to the same
  panels of Figure \ref{fig:wnsamp}.  The two density fields are strikingly
  different because long wavelengths have been suppressed in the right image
  by subtraction of coarse-cell means in Figure \ref{fig:wnsamp}.  The long
  wavelength components will be restored to the right image by addition of
  the coarse grid sample.}
  \label{fig:delsub}
\end{figure}

Figure \ref{fig:delsub} shows the result of convolving the two noise samples
of Figure \ref{fig:wnsamp} with the density transfer function of Figure
\ref{fig:transf}.  The left-hand panel gives $\xi_1*T$ while the right-hand
panel gives the desired short-wavelength field $\delta_1(\vec m,\vec n\,)$.
The two fields differ in the upper left quadrant because of the subtraction
of coarse-cell means from the white noise field used to generate the left
image.  Although the effect of this subtraction is barely evident in Figure
\ref{fig:wnsamp}, it dominates the comparison of the two panels in Figure
\ref{fig:delsub} because convolution by the transfer function acts as a
low-pass filter.  The left panel of Figure \ref{fig:delsub} gives a complete
sample of $\delta(\vec x\,)$ on a periodic grid of size 64 Mpc, while the
right panel shows only the short-wavelength components coming from mesh
refinement.

Careful examination of the right panel of Figure \ref{fig:delsub}
shows that the finite width of the transfer function has caused a little
smearing at the boundaries, which are matched periodically to the opposite
side of the box by the Fourier convolution.  (A few pixels along the right
and bottom edges of the left panel differ from green.)  However, these edge
effects do not represent errors in the short-wavelength density field.
Instead, they illustrate the fact that {\it outside} of the refined
region, the gravity field should include tidal contributions from the
short-wavelength fluctuations inside the refinement volume.  For the
purpose of computing $\delta_1(\vec m,\vec n\,)$ within the subvolume,
we simply discard everything outside the upper left quadrant.

As a test of our transfer function methods, we calculated $\delta_1(\vec m,
\vec n\,)$ using the exact transfer function instead of the spherical one.
The rms difference between the fields so computed was 0.0014 standard
deviations, a negligible difference.  As a test of the whole procedure,
we computed the power spectrum of the left panel of Figure \ref{fig:delsub}
and checked that it agrees within cosmic variance with the input
$\Lambda$CDM power spectrum.

We also compared the displacement field computed using the exact transfer
with that computed using the spherical one.  The rms difference was 0.0062
standard deviations, still negligible.  Then we compared the divergence
of the displacement field (computed in Fourier space as $i\vec k\cdot
\vec\psi$) with the density field, expecting them to agree perfectly.
Interestingly, this is not the case for the exact (or spherical) transfer
functions.  When the transfer functions are truncated in real space and
made periodic on a grid of size $(2M_s/M)L$, $T'_\delta(\vec k^{\,\prime\,})
\ne i\vec k^{\,\prime}\cdot\vec T'_\psi(\vec k^{\,\prime\,})$ despite the
fact that on the full refined grid $T_\delta(\vec k\,)=i\vec k\cdot\vec
T_\psi(\vec k\,)$.  The prime on the transfer functions indicates a {\it
different} Fourier space, as discussed after equation (\ref{congen4}).
The only way to test $-\vec\nabla\cdot\vec \psi=\delta$ for the exact
transfer function is to perform an FFT on the full refined grid of
$(rM)^3=1024^3$ grid points.  In \S \ref{sec:test} we will perform an
equivalent test with an end-to-end test of the entire method using a
$1024^3$ grid.

\subsection{Long Wavelength Components}
\label{sec:long}

Now we consider $\tilde\delta_0(\vec m,\vec n\,)$, the contribution to
the density field from the coarse grid.  As we see from equation
(\ref{del12a}), in principle we can compute $\tilde\delta_0$ by
spreading the original coarse-grid white noise sample $\xi_0(\vec m\,)$
to the $r^3$ subgrid points for each coarse grid point (i.e., all
$r^3M^3$ points) and then convolving with the high-resolution transfer
function as in equation
(\ref{congen2}):
\begin{equation}
  \label{tildel1}
  \tilde\delta_0(\vec m,\vec n\,)=\sum_{\vec m',\vec n^{\,\prime}}\xi_0
    (\vec m^{\,\prime})T(\vec m-\vec m^{\,\prime},\vec n-\vec n^{\,
    \prime})\ .
\end{equation}
However, this method is impractical, because the contributions to
$\tilde\delta_0$ coming from large distances are not negligible because
of the long range of the transfer functions (especially for the velocity
transfer function).  For the short wavelength field $\delta_1(\vec m,
\vec n\,)$ this causes no problems because the noise field $\xi_1$ is
nonzero only within the subvolume.  Here, however, the noise field
$\xi_0(\vec m\,)$ is nonzero over the entire simulation volume.
Including all relevant contributions in the convolution as written
would require working with the transfer function on the full grid of
size $(rM)^3$, which is exactly what we are trying to avoid.

A practical solution is to rewrite equation (\ref{tildel1}) as a
convolution of the coarse-grid density field with a short-ranged filter:
\begin{equation}
  \label{tildel2}
  \tilde\delta_0(\vec m,\vec n\,)=\sum_{\vec m',\vec n^{\,\prime}}
    \delta_0(\vec m^{\,\prime})W(\vec m-\vec m^{\,\prime},\vec n-
    \vec n^{\,\prime})\ .
\end{equation}
One can easily check that this is exactly equivalent to equation
(\ref{tildel1}) provided that
\begin{equation}
  \label{filtad}
  W(\vec m,\vec n\,)=(rM)^{-3}\sum_{\vec k}\exp[i\vec k\cdot\vec x(\vec m,
    \vec n\,)]\,{T(k)\over T(k_0)}
\end{equation}
where $\vec k_0=(2\pi/L)\vec\kappa$ is the projection of $\vec k$ into the
fundamental Brillouin zone (eq. \ref{kgrid}).

An exact evaluation of equation (\ref{tildel2}) still requires using a
full $(rM)^3$ grid.  However, we will see that $W(\vec x\,)$ falls off
sufficiently rapidly with distance that contributions to $\tilde\delta_0
(\vec m,\vec n\,)$ coming from large distances are negligible.  (This
will not be true for the velocity field, but we will develop a variation
to handle that case later.)  Thus, we may truncate $W(\vec x\,)$ at the
boundary of the refinement region and perform the convolution of equation
(\ref{tildel2}) using a $(2rM_s)^3$ grid just as we did for the
short-wavelength field.  The errors of this procedure will be quantified
below.

Equation (\ref{tildel2}) has a simple interpretation.  The coarse-grid
density field $\delta_0(\vec m\,)$ is spread to the fine grid by replicating
the coarse-grid values to each of the $r^3$ grid points within
a single coarse grid cell.  The result is an artifact called aliasing.
In real space this artifact is manifested by having constant values
within pixels larger than the spatial resolution.  In $k$-space the
effect is to replicate low-frequency power in the fundamental
Brillouin zone to higher frequencies.  Thus, the wrong transfer function
is used if one simply sets $\tilde\delta_0(\vec m,\vec n\,)$ to $\delta_0
(\vec m\,)$.  Equation (\ref{filtad}) defines an anti-aliasing filter
which corrects the transfer function from the coarse grid (with wavevectors
$\vec k_0$ in the fundamental Brillouin zone) to the full $k$-space.
It smooths the sharp edges that arise from spreading $\delta_0(\vec m\,)$
to the fine grid.  The anti-aliasing filter removes the artifacts caused
by replication of the fundamental Brillouin zone.

The anti-aliasing filter is manifestly nonspherical, so we cannot use the
spherical transform method of \S \ref{sec:spherical} to evaluate it.
However, $W(\vec m,\vec n\,)$ is sharply peaked.  This is obvious from
the fact that its Fourier transform is constant over the fundamental
Brillouin zone; the Fourier transform of a constant is a delta function.
Thus, we expect $W(\vec m,\vec n\,)$ to be peaked on the scale of a few
coarse grid spacings.  As a result, the minimal $k$-space sampling method
of \S \ref{sec:minimal} should suffice (with a variation for the velocity
field).

The division by $T(k_0)$ in equation (\ref{filtad}) requires that we
compute the coarse-grid density field $\delta_0(\vec m\,)$ without a
Hanning filter; otherwise $T(k_0)$ would be zero in the corners of each
Brillouin zone.  Simply put, if we want to correctly sample the density
field at high resolution, we should not cut out long-wavelength power
by filtering.  However, we will apply a spherical Hanning filter at
the shortest wavelength to remove the anisotropic structure that was
apparent in Figure \ref{fig:transd3}.

At the corners of each Brillouin zone $k_0=0$ but $k\ne0$ (aside from
the fundamental mode for the whole box).  At these wavevectors,
$\delta_0(\vec m)$ has no power and so no error is made by setting
$T(k)/T(k_0)\to0$.  In the case of the displacement field, each component
of $\vec T_\psi(\vec k_0)$ vanishes along an entire face of the Brillouin
zone, as explained in the paragraph before equation (\ref{gauss}).
We also set to zero these contributions to the Fourier series in equation
(\ref{filtad}).

\begin{figure}[t]
\includegraphics[scale=0.4]{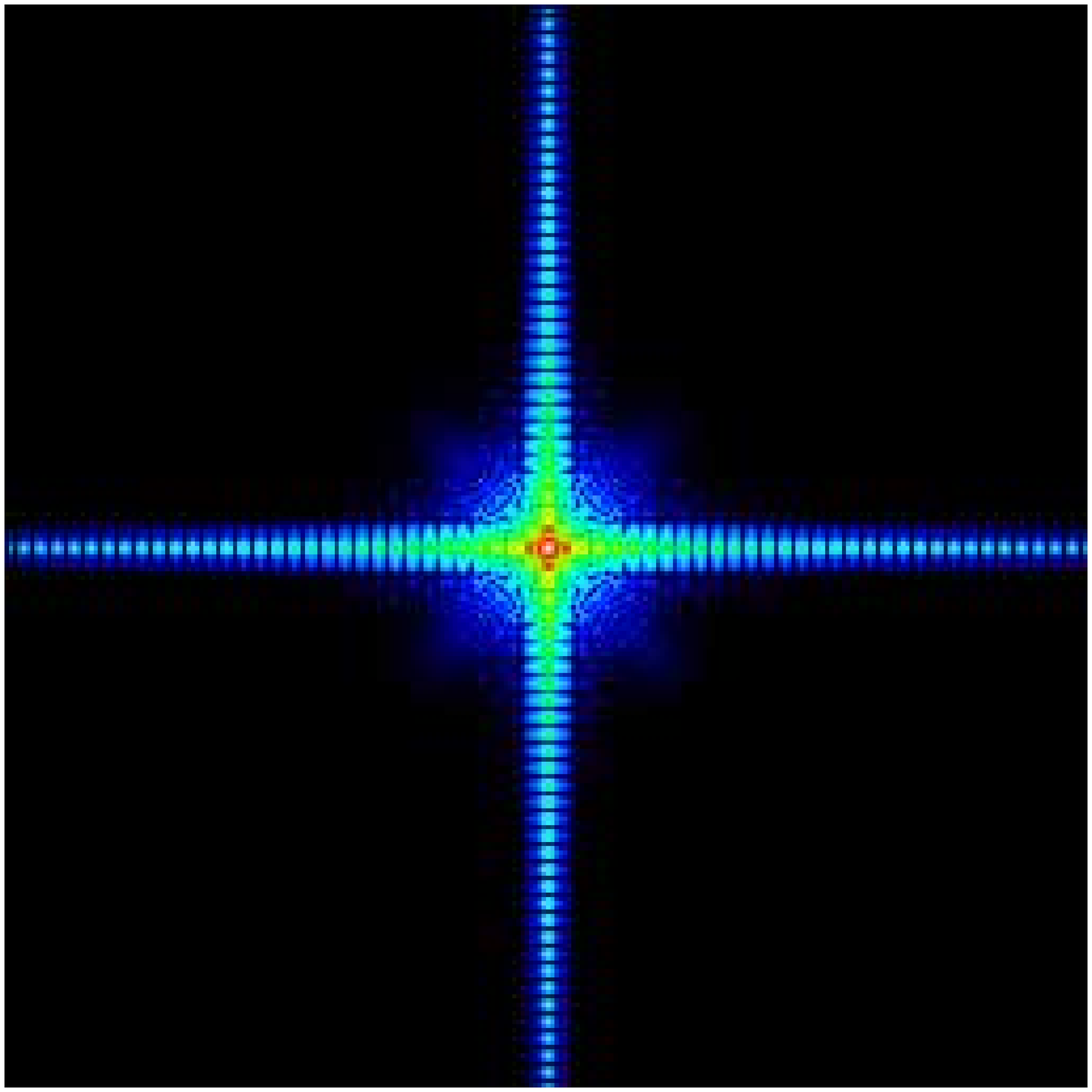}  
\includegraphics[scale=0.4]{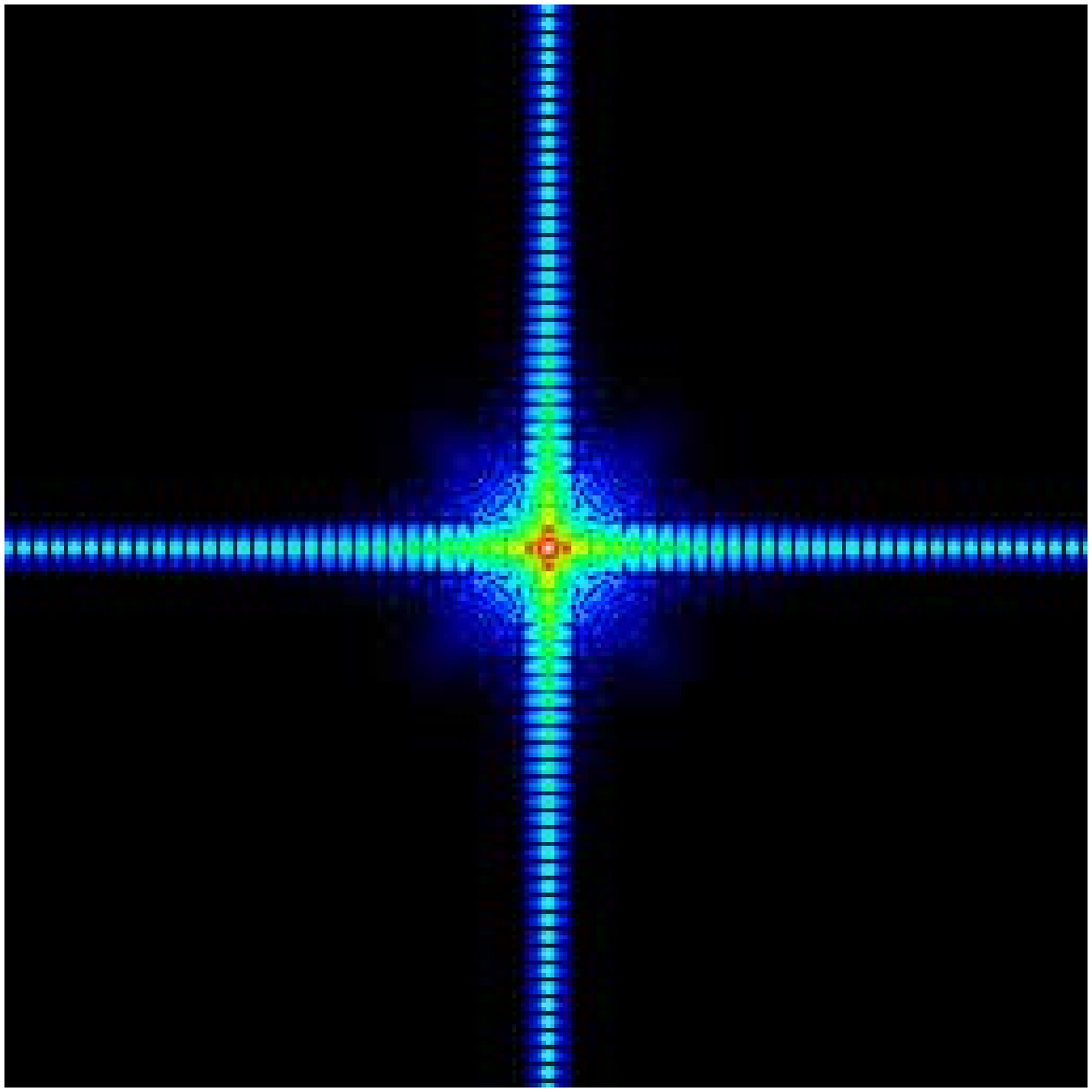}  
  \caption{Anti-aliasing filter $W(\vec m,\vec n\,)$ for the density.
  The left panel shows the filter computed using the exact method of
  \S \ref{sec:exact} while the right panel uses the minimal $k$-space
  sampling method of \S \ref{sec:minimal}. False colors are scaled to
  the logarithm of absolute value of the filter, which shows six orders
  of magnitude.  The banded appearance is caused by low-amplitude
  oscillations. The oscillations act to smooth the sharp edges of the
  coarse grid fields when they are refined to the subgrid.  The difference
  between the two filters is negligible away from the edges.}
  \label{fig:filtd}
\end{figure}

Figure \ref{fig:filtd} shows the density anti-aliasing filter computed
with a spherical Hanning filter for $T(k)$.  The minimal $k$-space method
gives good agreement with the much slower exact calculation.  Along the
axes at the edges of the volume the errors are up to a factor of two,
but $W$ is very small and oscillates, making these errors unimportant.
The banding is due to the sign oscillations of $W$.  They have a
characteristic scale equal to the coarse grid spacing and they arise
because of the discontinuity of $T(k)/T(k_0)$ at Brillouin zone boundaries
in equation (\ref{filtad}).  Such oscillations are characteristic of
anti-aliasing filters.  The filter falls off sufficiently rapidly with
distance from the center that we can expect accurate results by truncating
it outside the region shown.

\begin{figure}[t]
\includegraphics[scale=0.4]{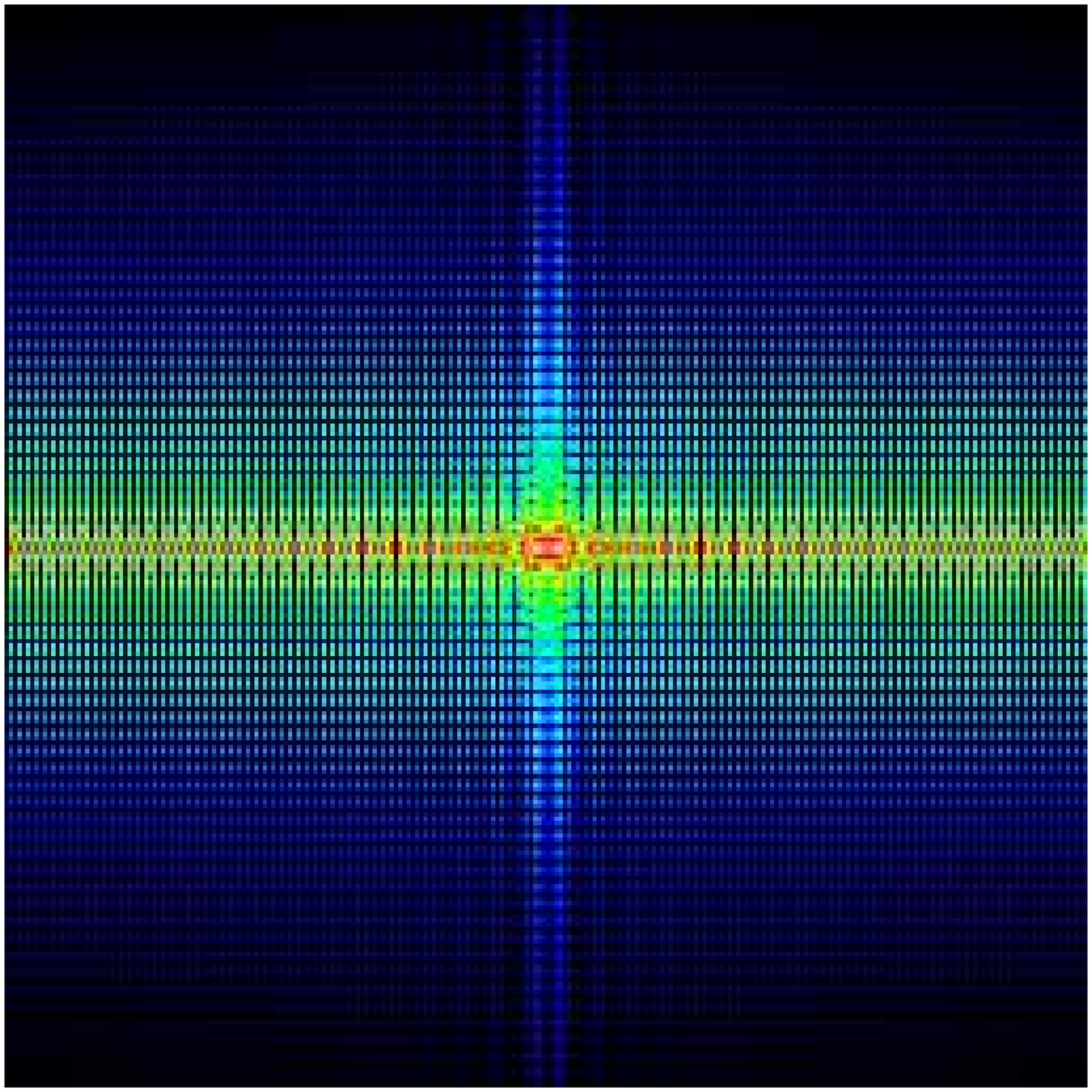}  
\includegraphics[scale=0.4]{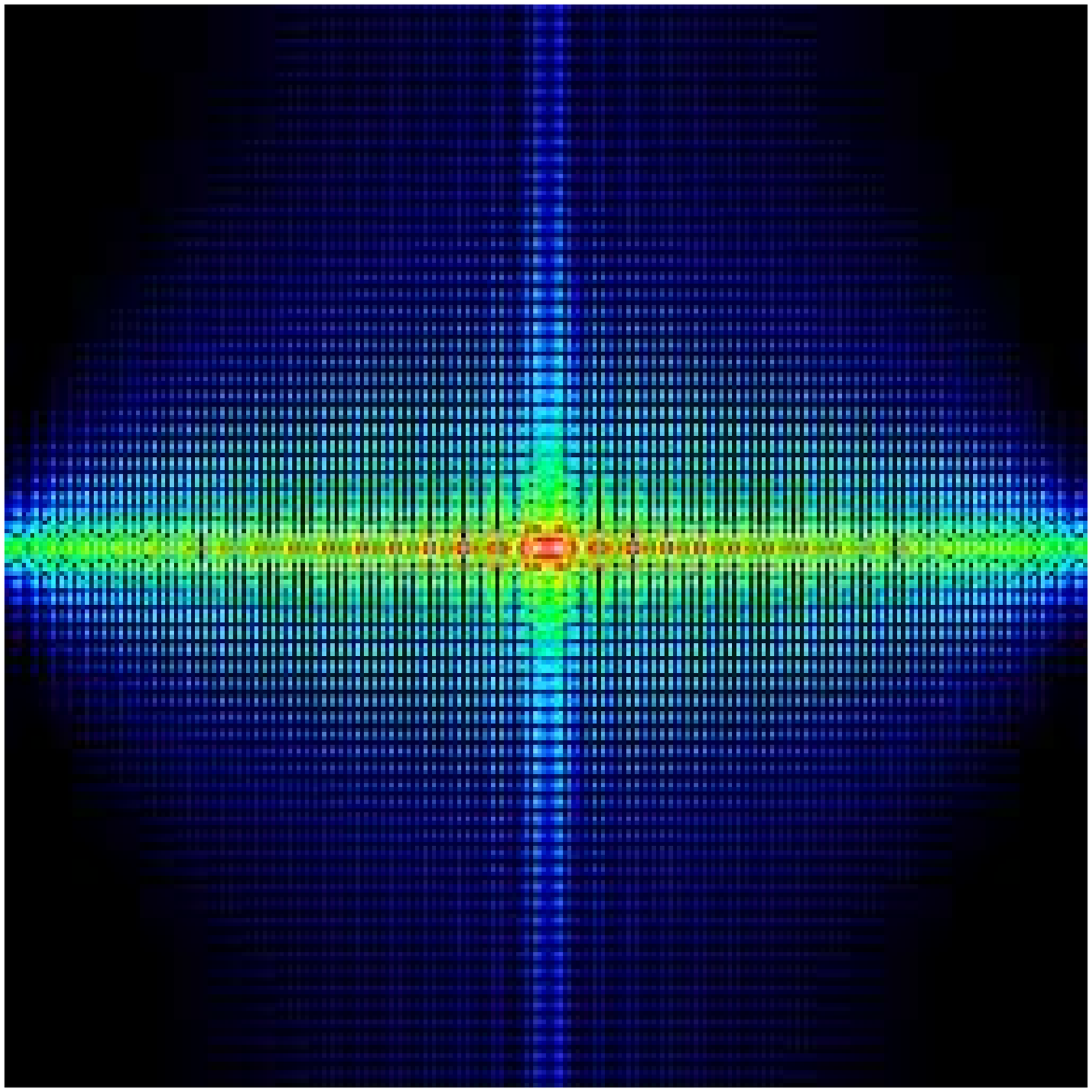}  
  \caption{Anti-aliasing filters for the $x$-component of displacement
  (or velocity or gravity), computed with the exact (left) and minimal
  $k$-space sampling (right) methods.  False colors are scaled to the
  logarithm of absolute value of the filter spanning four orders of
  magnitude. Because the displacement is sensitive to longer wavelengths
  than the density, the differences between the two computational methods
  here is more pronounced than for the density filter of Figure
  \ref{fig:filtd}.}
  \label{fig:filtx}
\end{figure}

Figure \ref{fig:filtx} shows the corresponding result for the displacement
(or velocity or gravity) field filter.  (Recall that the displacement,
velocity, and gravity are proportional to one another in linear theory.)  Now
the errors of the minimal $k$-space sampling method are significant.  They
arise because the minimal sampling method forces $W(\vec m,\vec n\,)$ to
be periodic on the scale of the box shown in the figure (twice the subgrid
size) while with the exact method the scale of periodicity is larger by a
factor $M/2M_s$ (or 4 in this case).  In other words, the filter does not
fall off very rapidly with distance, so truncating it and making it periodic
in the box of size $2rM_s$ introduces noticeable errors.  However, because
the filter is still sharply peaked and oscillatory with small amplitude, it
is possible that these errors are negligible.  We will quantify the errors
in \S \ref{sec:test}.

The procedure is now similar to that of \S \ref{sec:short}.  Once we have
the anti-aliasing filters, the next step is to obtain samples of the
coarse-grid fields $\delta_0(\vec m\,)$ and $\vec\psi_0(\vec m\,)$ that
we wish to refine.  We do this using the convolution method of \S
\ref{sec:disconv}.  For testing purposes, we construct a coarse-grid sample
of white noise, $\xi_0(\vec m\,)$, which exactly equals the long-wavelength
parts of the noise shown in the left panel of Figure \ref{fig:wnsamp}.
This was achieved by modifying {\tt GRAFIC} \citep{b95} to sample a white
noise field $\xi(\vec m\,)$ in the spatial domain.  Fourier transformation
to $\xi(\vec k\,)$ then allows the calculation of density (and similarly
displacement) by equation (\ref{delfft}).  As a result, we have chosen
our coarse grid sample so that $\xi_0(\vec m\,)=\bar\xi_1(\vec m\,)$ within
the refinement subgrid.  This choice is made so that later we can see
directly how long- and short-wavelength components of the noise contribute
to the final density field.

\begin{figure}[t]
\includegraphics[scale=0.4]{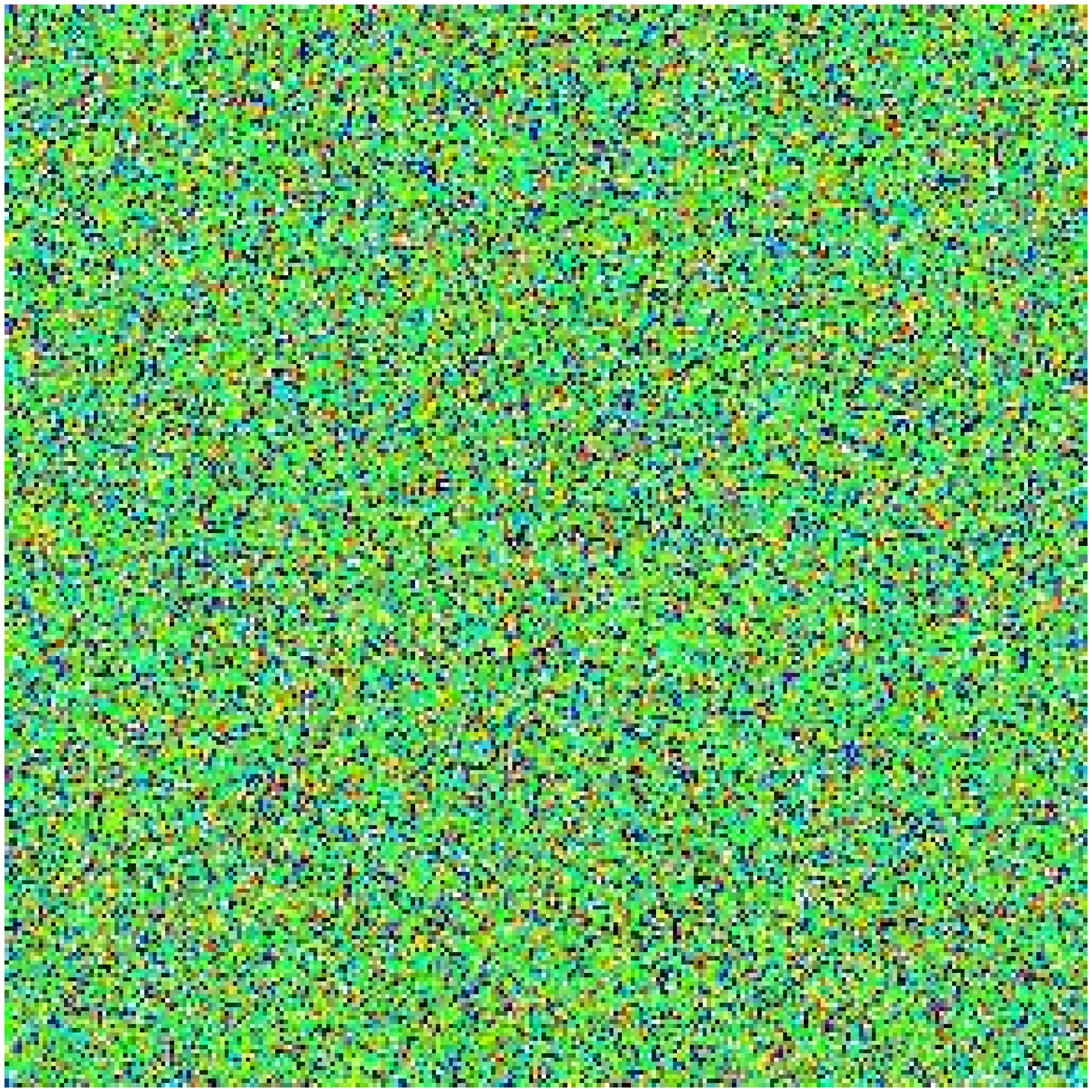}  
\includegraphics[scale=0.4]{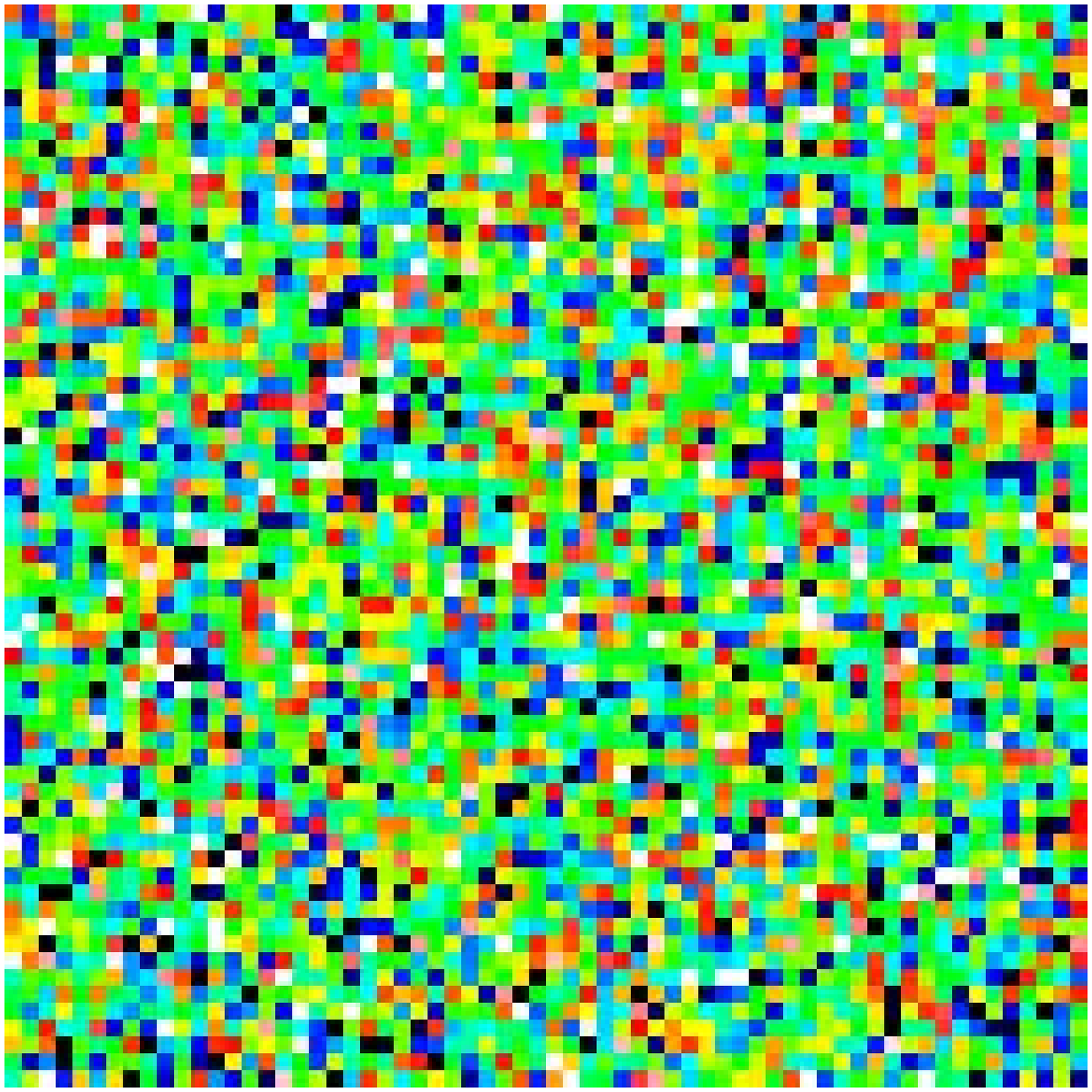}    
  \caption{Slices through the white noise sample $\xi_0(\vec m\,)$ for the
  coarse grid. Left: Full cube of size 256 Mpc.  Right: Magnification of the
  upper left corner by a factor of 4 to show the region that will be refined.
  Aliasing (sharp pixel boundaries) is now evident.  False colors
  are scaled to linear values ranging from $\pm2$ standard deviations.}
  \label{fig:cnsamp}
\end{figure}

Figure \ref{fig:cnsamp} shows the white noise sample adopted for the coarse
grid.  The right panel is obtained by averaging the left panel of
Figure \ref{fig:wnsamp} over $4^3$ subgrid mesh points (1 Mpc$^3$ volume).
Figure \ref{fig:wnsamp} shows only a single thin slice of width 0.25 Mpc
while Figure \ref{fig:cnsamp} shows coarse cells of thickness 1 Mpc, so one
should not expect the two figures to appear similar.  The left panel
of Figure \ref{fig:cnsamp} shows a full slice of size 256 Mpc, obtained by
filling out the rest of the volume with white noise.

\begin{figure}[t]
\includegraphics[scale=0.4]{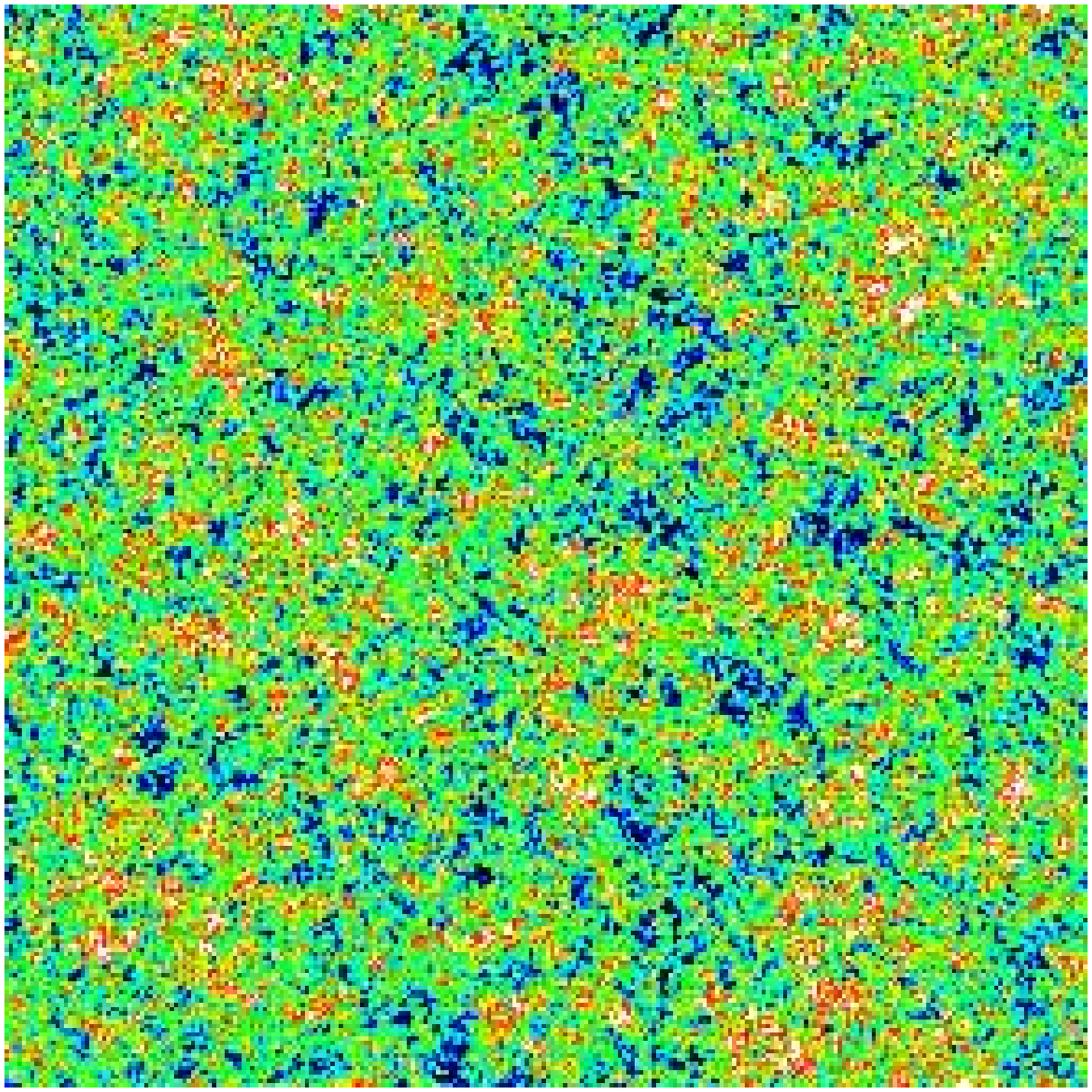}  
\includegraphics[scale=0.4]{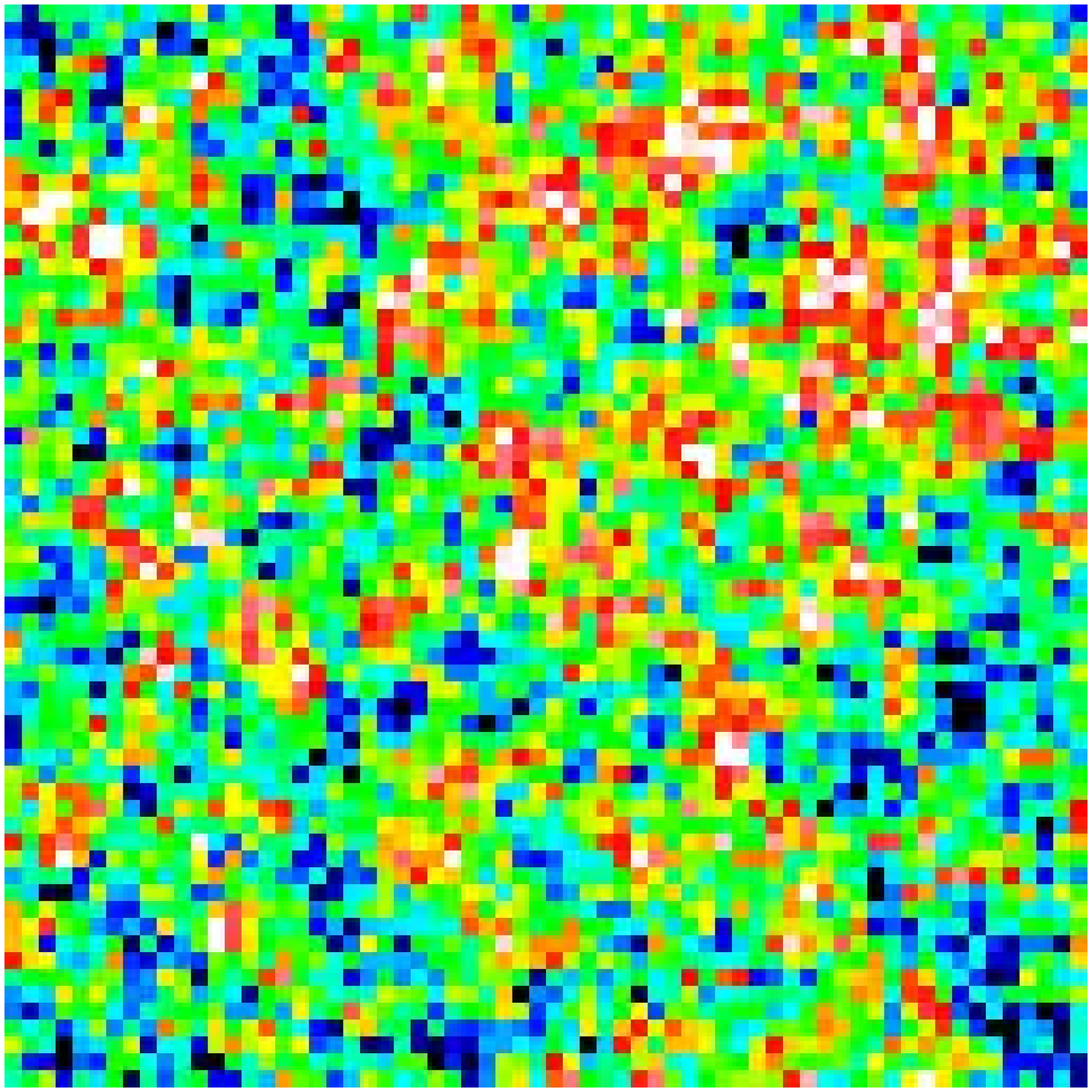}  
  \caption{Slices through the coarse grid density field $\delta_0(\vec m\,)$
  resulting from convolution of Figure \ref{fig:cnsamp} with the density
  transfer function sampled on the coarse grid.
  Left: Full cube of size 256 Mpc, with periodic boundary conditions.
  Right: Magnification of the upper left corner by a factor of 4 to show
  a square of size 64 Mpc.  This panel shows the 32 Mpc region that we wish
  to refine to include the correct small-scale power.  The magnified pixels
  represents an aliasing artifact.  False colors are scaled as in Figure
  \ref{fig:delsub}.  Random numbers were chosen so that the right panel
  corresponds to the coarsely sampled long wavelength components of the
  left panel of Figure \ref{fig:delsub}.}
  \label{fig:deltop}
\end{figure}

This white noise sample on the coarse grid was convolved with the transfer
function using {\tt GRAFIC} to give the coarse density field $\delta_0
(\vec m\,)$ that we wish to refine.  The results are shown in Figure
\ref{fig:deltop}.  The right panel shows a 64 Mpc subvolume including the
32 Mpc refinement region.  The obvious pixelization is the result of mesh
refinement: the coarse grid density field has been spread to the fine grid.
This pixelization causes power from wavelengths longer than the coarse grid
spacing to be aliased to higher frequencies.  If uncorrected, this aliasing
would introduce spurious features into the power spectrum.  Thus, the coarse
grid sample must be convolved with an anti-aliasing filter as described
above.

\begin{figure}[t]
\includegraphics[scale=0.4]{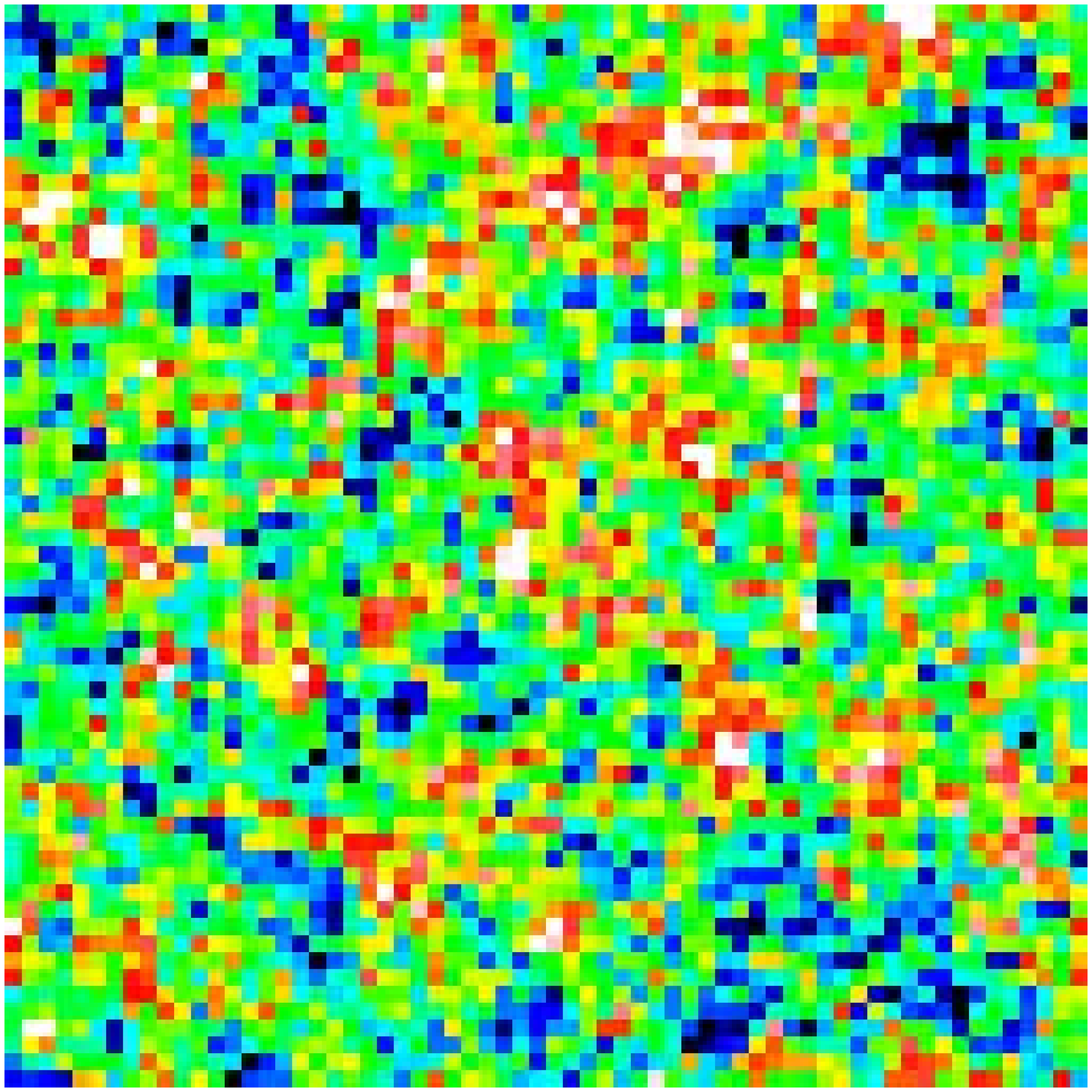}
\includegraphics[scale=0.4]{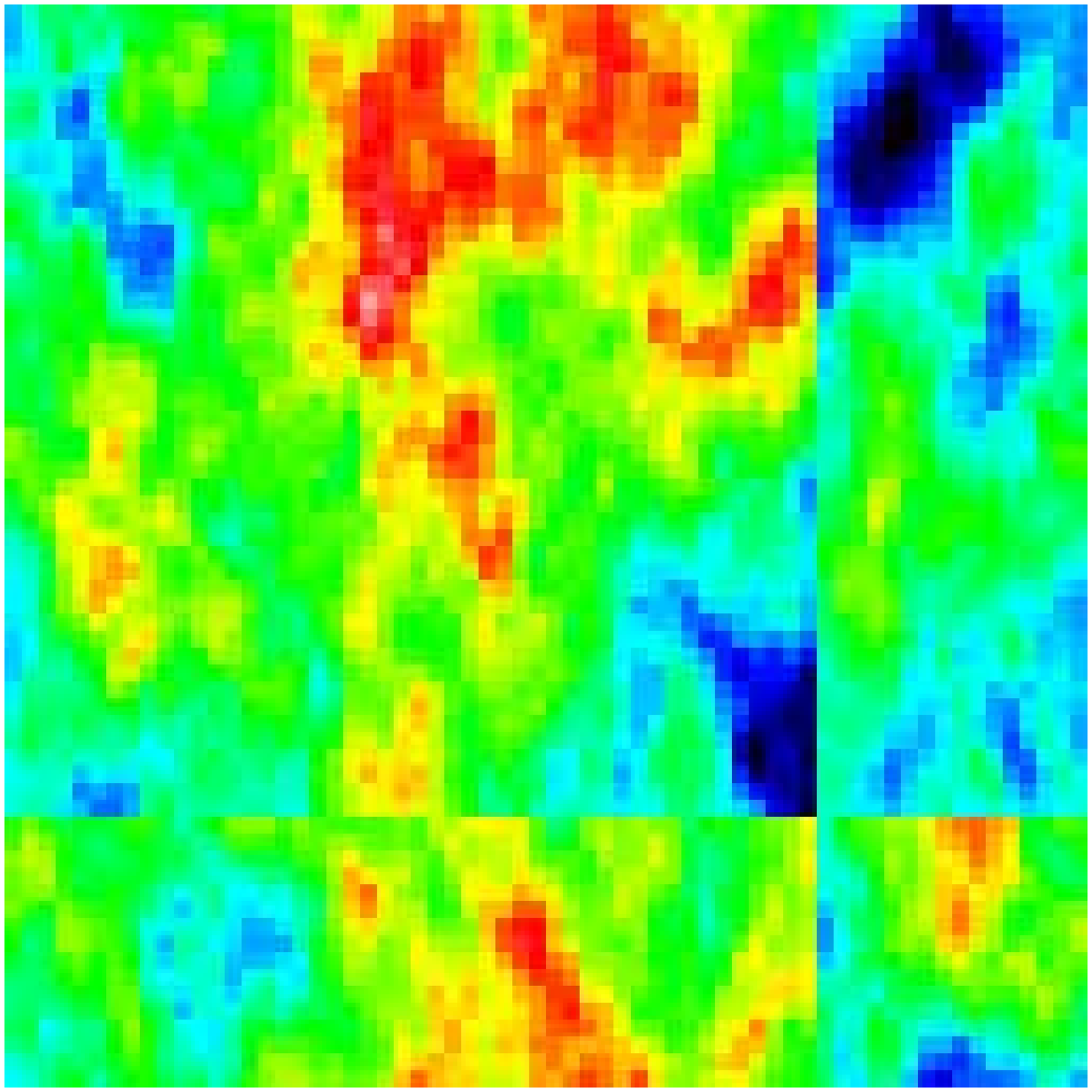}
  \caption{Slices through the density (left) and velocity (x-component,
  right) fields on the coarse grid after a buffer region of width 16 Mpc
  has been placed around the 32 Mpc subvolume in the upper left.  In the
  upper left 48 Mpc area, the left panel matches the right panel of Figure
  \ref{fig:deltop}.  The bottom and right quartiles are filled with values
  from the top and left of the subvolume which were then wrapped
  periodically.  This is clearer for the velocity field because of its
  larger coherence length.  The buffer regions and periodic boundary
  conditions are needed because of the FFT-based method for convolution
  with anti-aliasing filters.}
  \label{fig:delvlong}
\end{figure}

Special care is needed with the boundary conditions for the anti-aliasing
convolution of equation (\ref{tildel2}).  The top grid density field
$\delta_0(\vec m\,)$ fills the subgrid shown in the right panel of Figure
\ref{fig:deltop} without periodic boundary conditions.  The anti-aliasing
filter (Fig. \ref{fig:filtd}) has finite extent; therefore FFT-based
convolution of the two will lead to spurious contributions to $\tilde
\delta_0$ at the subvolume edges coming from $\delta_0$ on the opposite
side of the box.  To avoid this, we surround the subvolume (which occupies
one octant of the convolution volume) with a buffer region of width
one-half of the subvolume in each dimension.  The correct density values
from the top grid are placed in this buffer.  Because our subvolume is not
centered but rather is placed in the corner of the cube of size $2rM_s$,
we wrap half of the buffer to the other side of this cube.  The results are
shown in Figure \ref{fig:delvlong}.

\begin{figure}[t]
\includegraphics[scale=0.4]{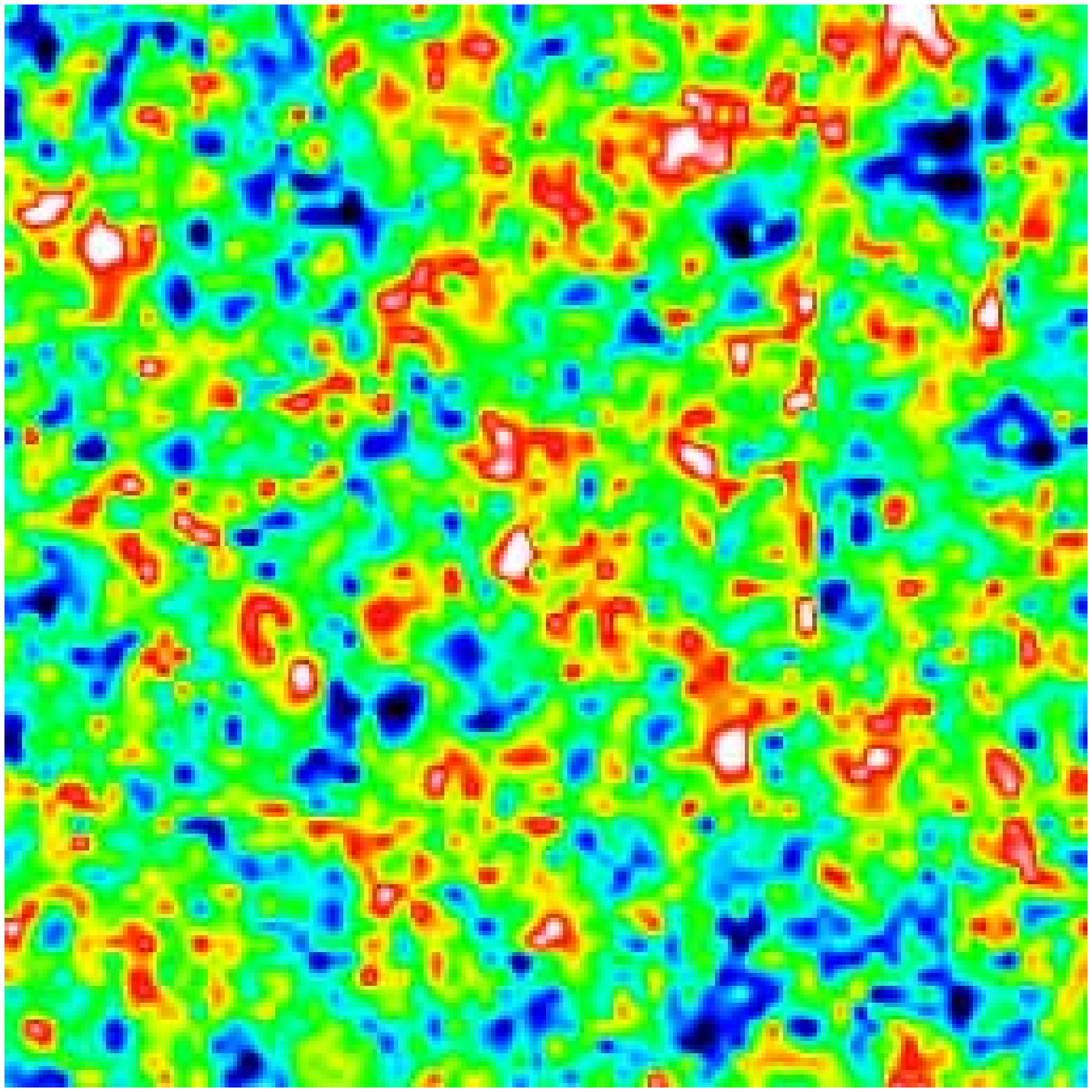}
\includegraphics[scale=0.4]{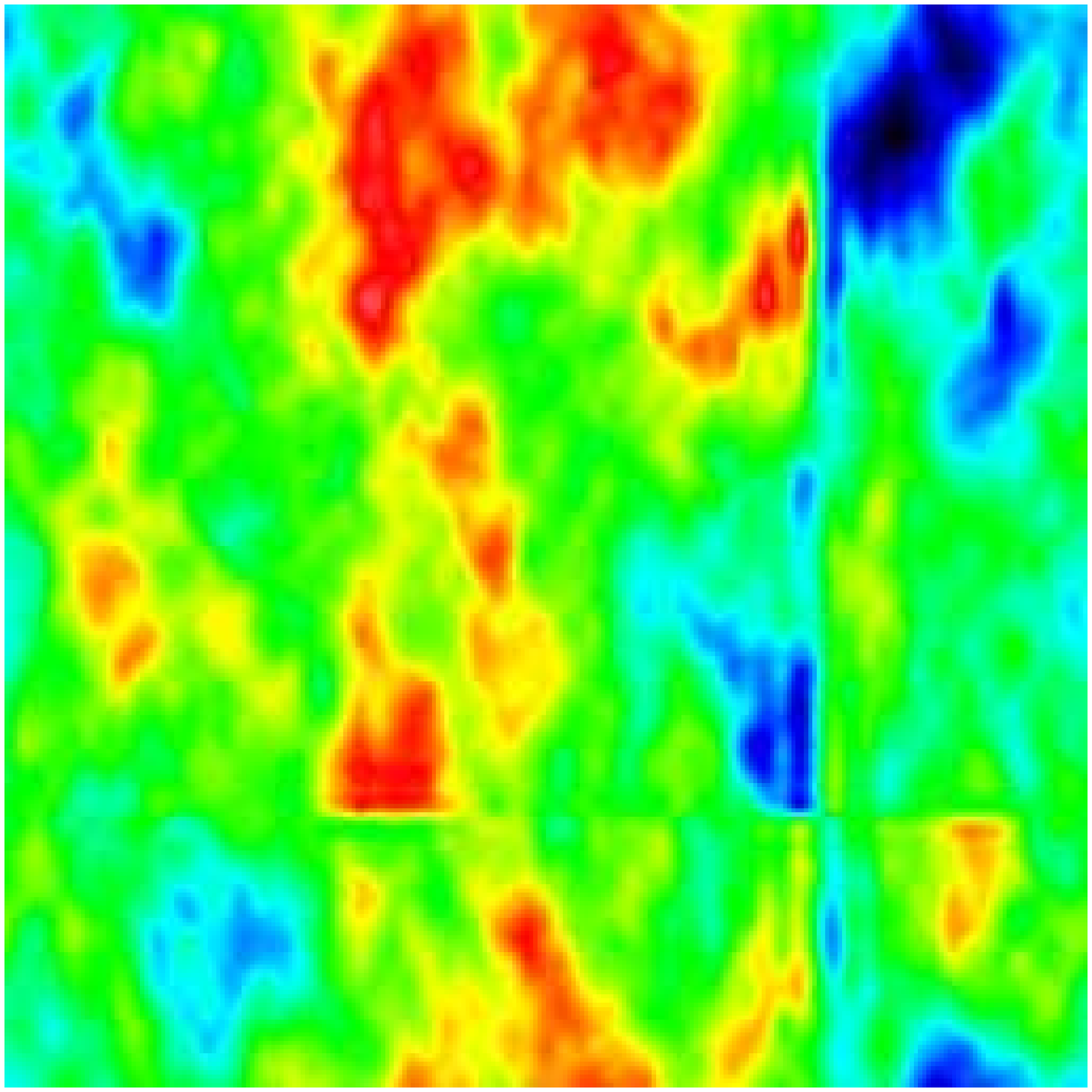}
  \caption{Long-wavelength density and velocity fields after convolution
  by the anti-aliasing filter.  Each panel is the convolution of the
  corresponding panel of Figure \ref{fig:delvlong} by the appropriate
  anti-aliasing filter (Fig. \ref{fig:filtd} for the density, Fig.
  \ref{fig:filtx} for the velocity).  The anti-aliasing filters have
  eliminated the pixelization artifacts present in Figure \ref{fig:delvlong}.
  Convolution across the discontinuity at the boundary of the buffer region
  causes some errors but these are small within the desired refinement
  region (the upper left quadrant in these images).}
  \label{fig:condel1}
\end{figure}

Figure \ref{fig:condel1} shows the density field $\tilde\delta_0(\vec m,
\vec n\,)$ and the corresponding velocity field after convolution with the
anti-aliasing filter $W$.  The minimal $k$-space filter has been used here;
there would be almost no discernible difference if the exact filter was
used instead.  The pixelated images of Figure \ref{fig:delvlong} have now
been smoothed appropriately for the transfer function.  Smoothing over
pixelization artifacts is the purpose behind anti-aliasing filters, whether
they be applied in image processing or cosmology.

The convolution method used here is not exact.  Quantifying its errors
requires evaluating equation (\ref{tildel1}) or (\ref{tildel2}) using a
full convolution of size $(rM)^3$.  We do this in the next subsection, where
we test all stages of the mesh refinement method.

\subsection{Testing the Refined Fields}
\label{sec:test}

\begin{figure}[t]
\includegraphics[scale=0.4]{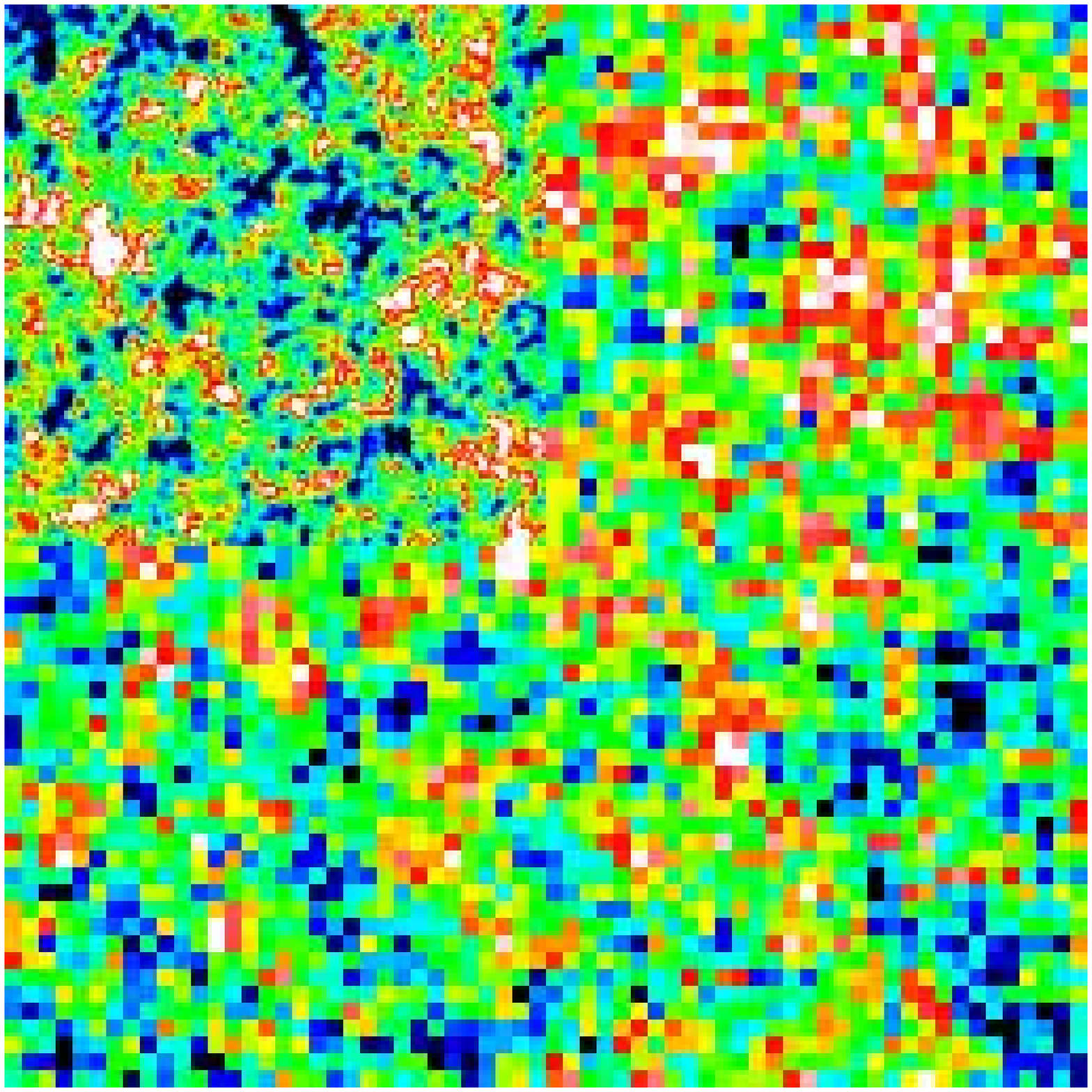}   
\includegraphics[scale=0.4]{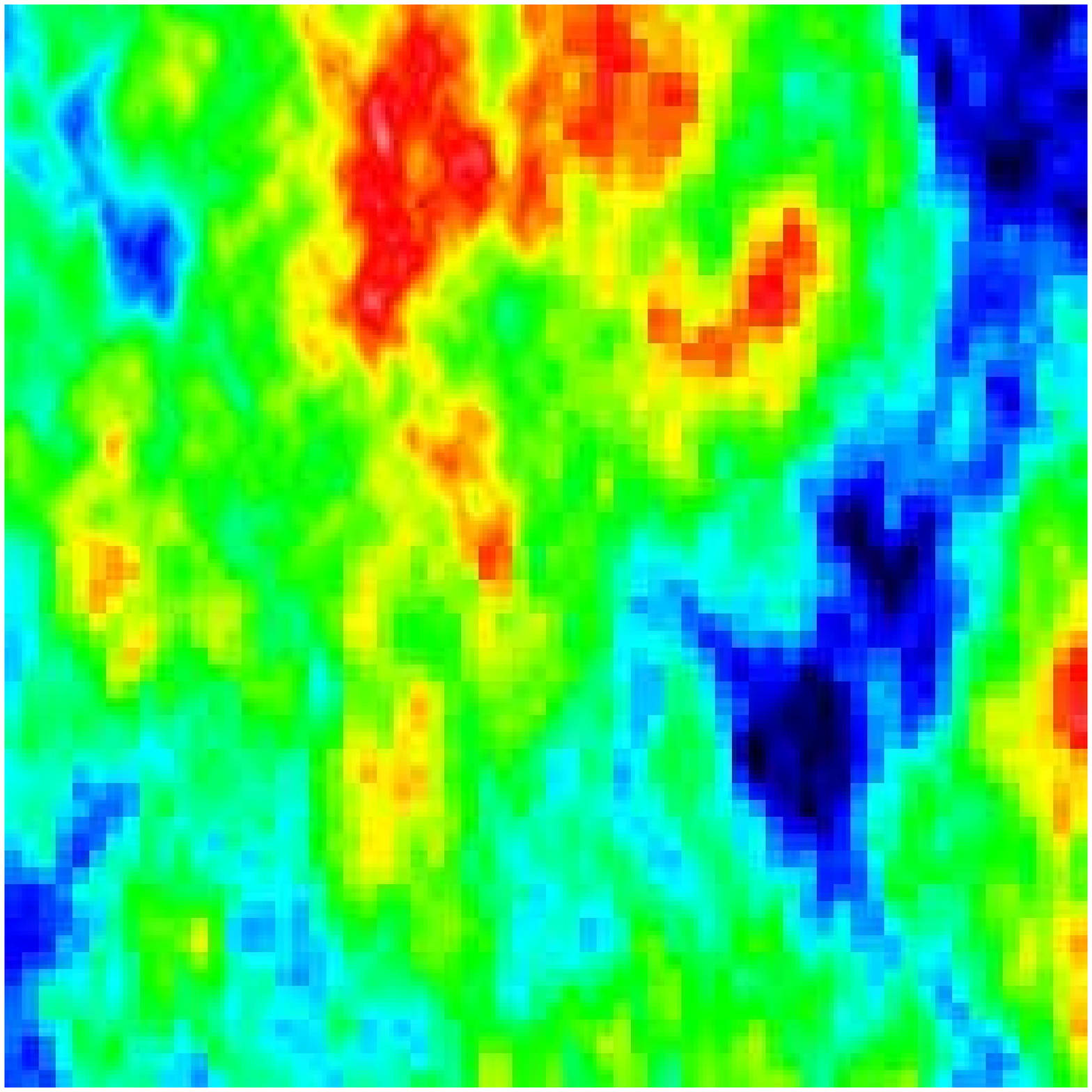}     
  \caption{Slices through the 2-level linear density (left) and velocity
  (right, $x$-component) fields in a region 64 Mpc across extracted from
  the 256 Mpc realization.  False colors are scaled to linear values
  ranging from $\pm2$ standard deviations of the high-resolution fields.
  The refinement subgrid is the upper left quadrant in each case.  Outside
  of this region the coarse (top) grid values are shown to illustrate how
  mesh refinement increases the resolution.  The density figure may be
  compared directly with the right-hand panel of Figure \ref{fig:deltop}.}
  \label{fig:multi2}
\end{figure}

Having computed separately the short- and long-wavelength contributions
to the density and velocity (or displacement) fields, we combine them in
Figure \ref{fig:multi2} using equation (\ref{del12}) to give the
complete multiscale fields.  The four-fold increase in resolution can be
seen by comparing the subvolume with the rest of the field.  The effects
of higher resolution are much more pronounced for the density than they
are for the velocity because of the density field's steeper dependence on
wavenumber.

A test of the entire mesh refinement procedure can be made by generating
the density and velocity fields at full $1024^3$ resolution over the whole
256 Mpc box.  This was done by modifying the author's {\tt GRAFIC} code
\citep{b95} to replace its random numbers in $k$-space with an input
white noise field in real space.  The noise field was constructed to match
the upper left quadrant of the left panel of Figure \ref{fig:wnsamp} in
a high-resolution region 32 Mpc across and to match the left panel of
Figure \ref{fig:cnsamp} everywhere else, with noise values made uniform
in 1 Mpc cells ($4^3$ grid cells of the $1024^3$ grid).  Thus, the white
noise field was sampled as in Figure \ref{fig:amr}.  In order to have
sufficient computer memory, {\tt GRAFIC} was run on the Origin 2000
supercomputer at the National Computational Science Alliance.

\begin{figure}[t]
\includegraphics[scale=0.4]{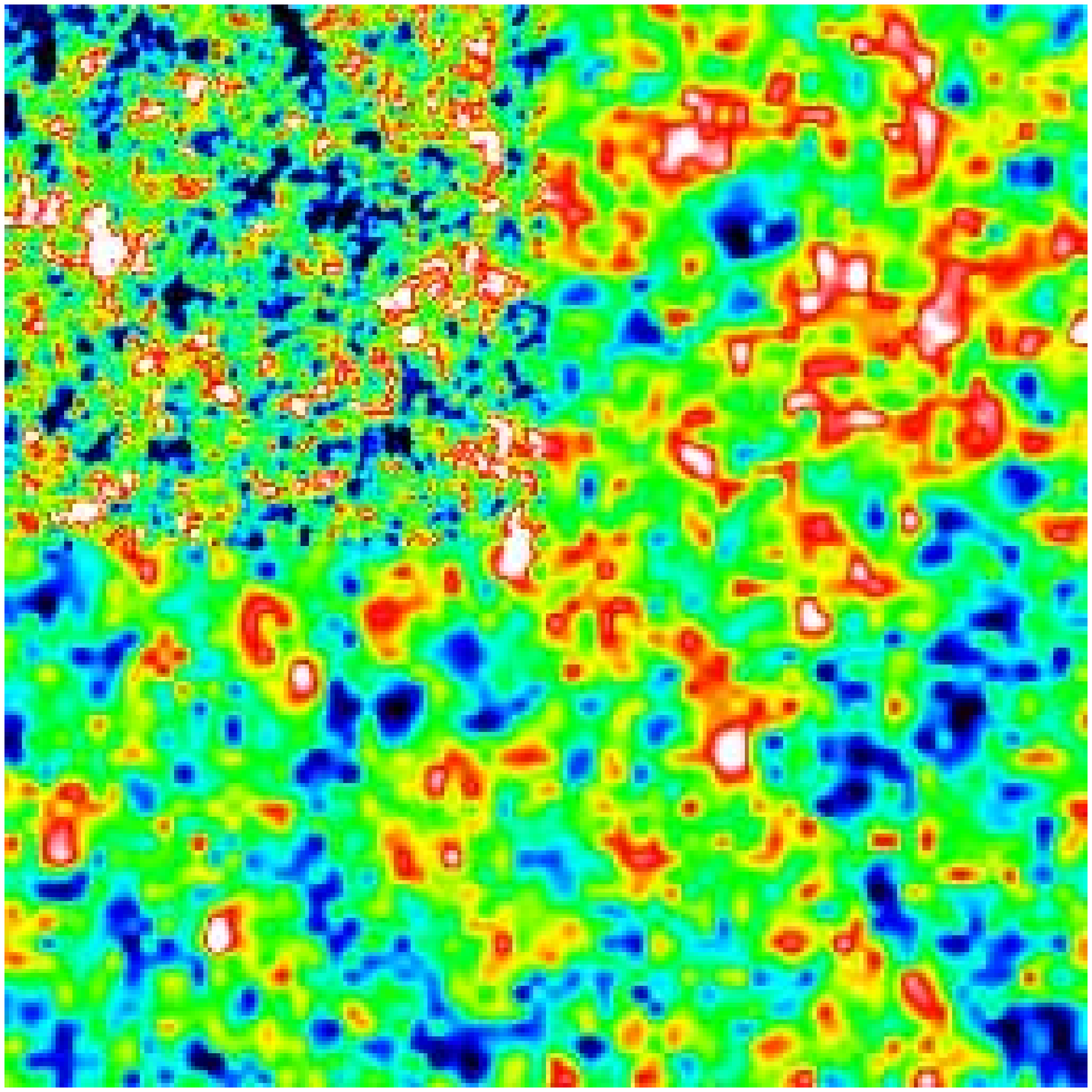}   
\includegraphics[scale=0.4]{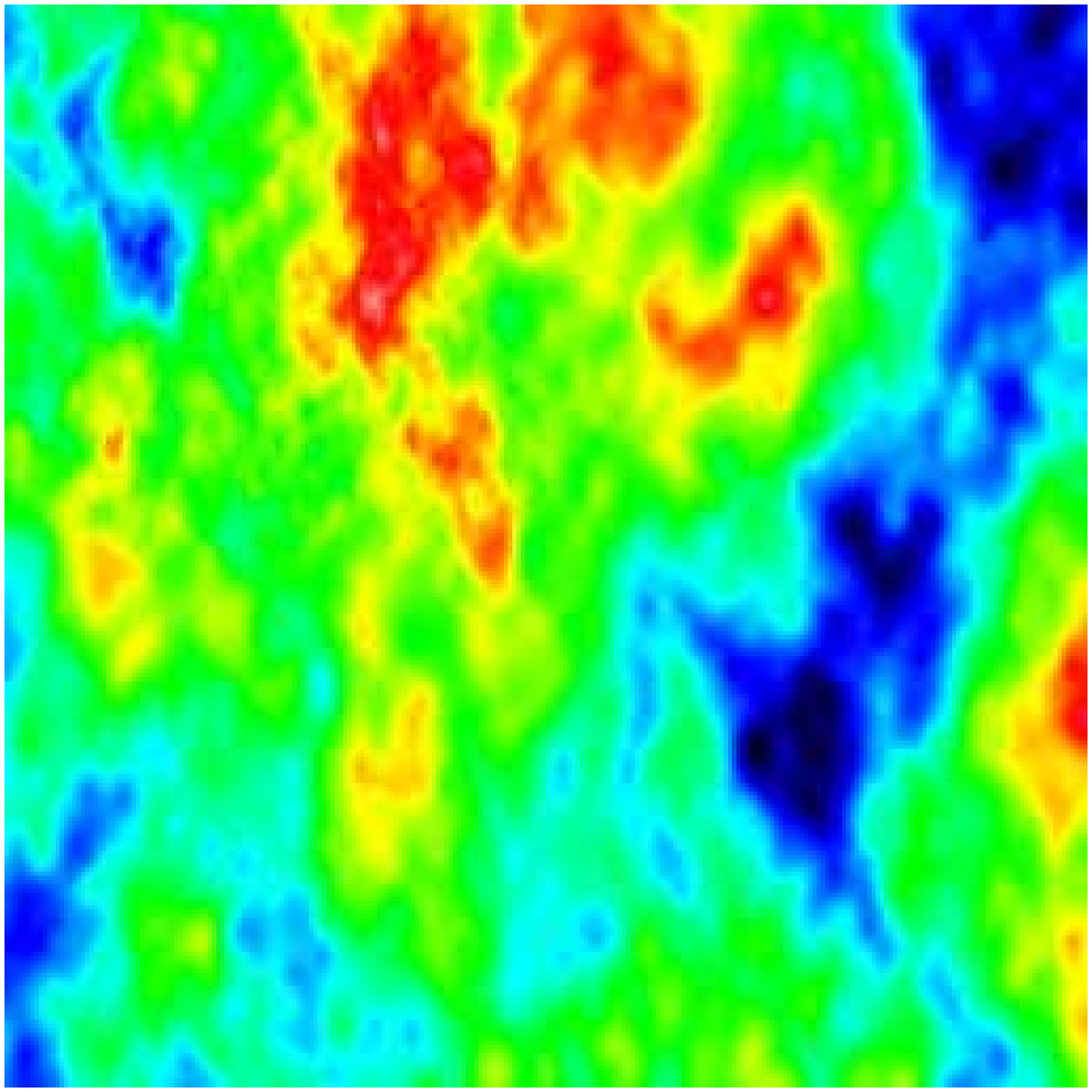}    
  \caption{Density (left) and velocity (right) fields computed using a full
  $1024^3$ grid with random numbers chosen to match the multiscale
  calculation.  This figure gives the exact results against which to
  compare Figure \ref{fig:multi2}.}
  \label{fig:grafic1024}
\end{figure}

The results of this full-resolution calculation are shown in Figure
\ref{fig:grafic1024}.  The high-resolution fields are smooth outside of
the refinement volume simply because they have been convolved with a
high-resolution transfer function; by contrast, Figure \ref{fig:multi2}
shows only the sampling of a low-resolution mesh outside of the subvolume.
These resolution differences are not important here.  Rather, it is the
comparison in the high-resolution subvolume that is important.  Evidently
the density field is accurately reproduced by the multiscale algorithm
while there are some visible errors in the velocity field.

\begin{figure}[t]
\includegraphics[scale=0.4]{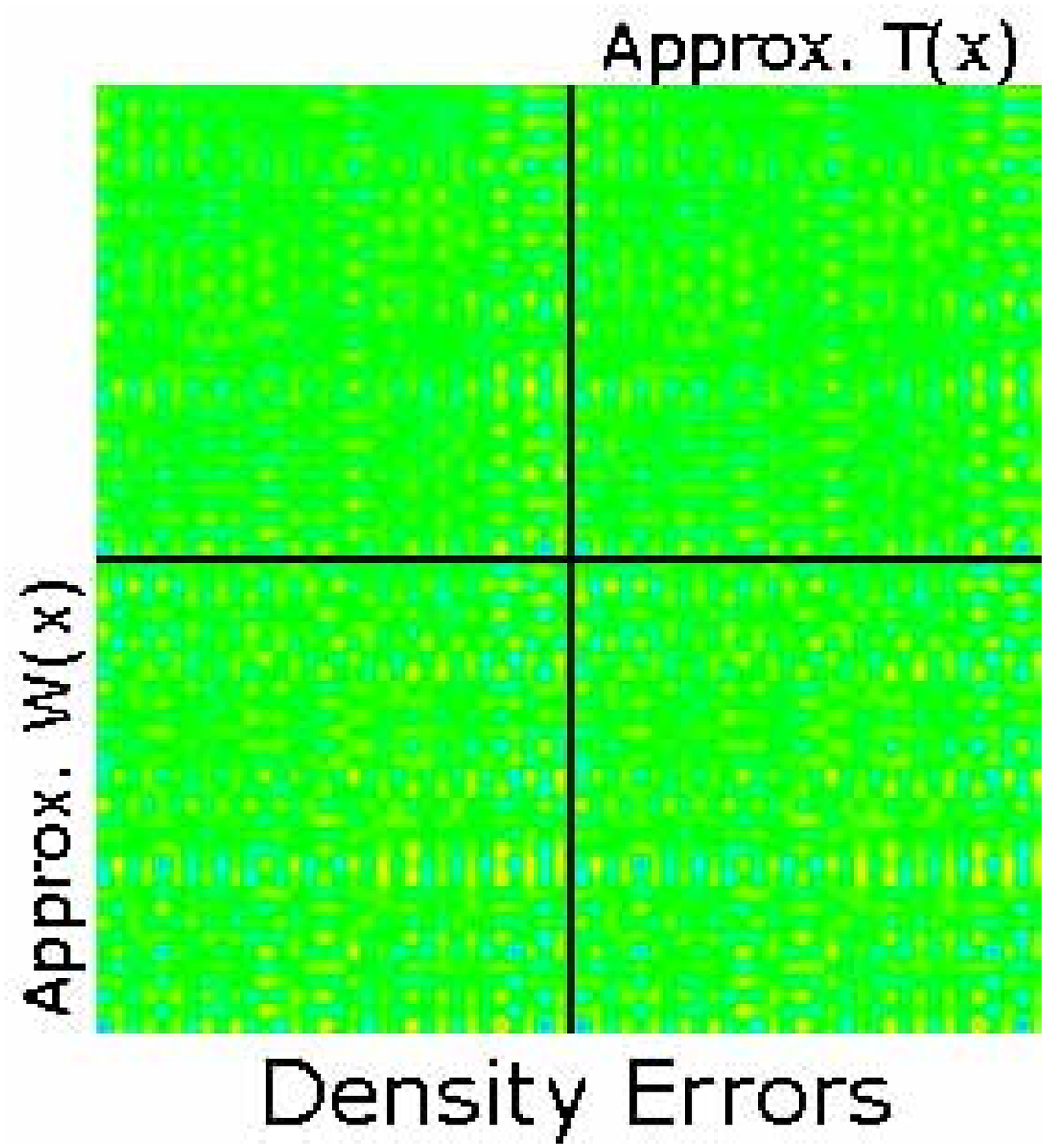}
\includegraphics[scale=0.4]{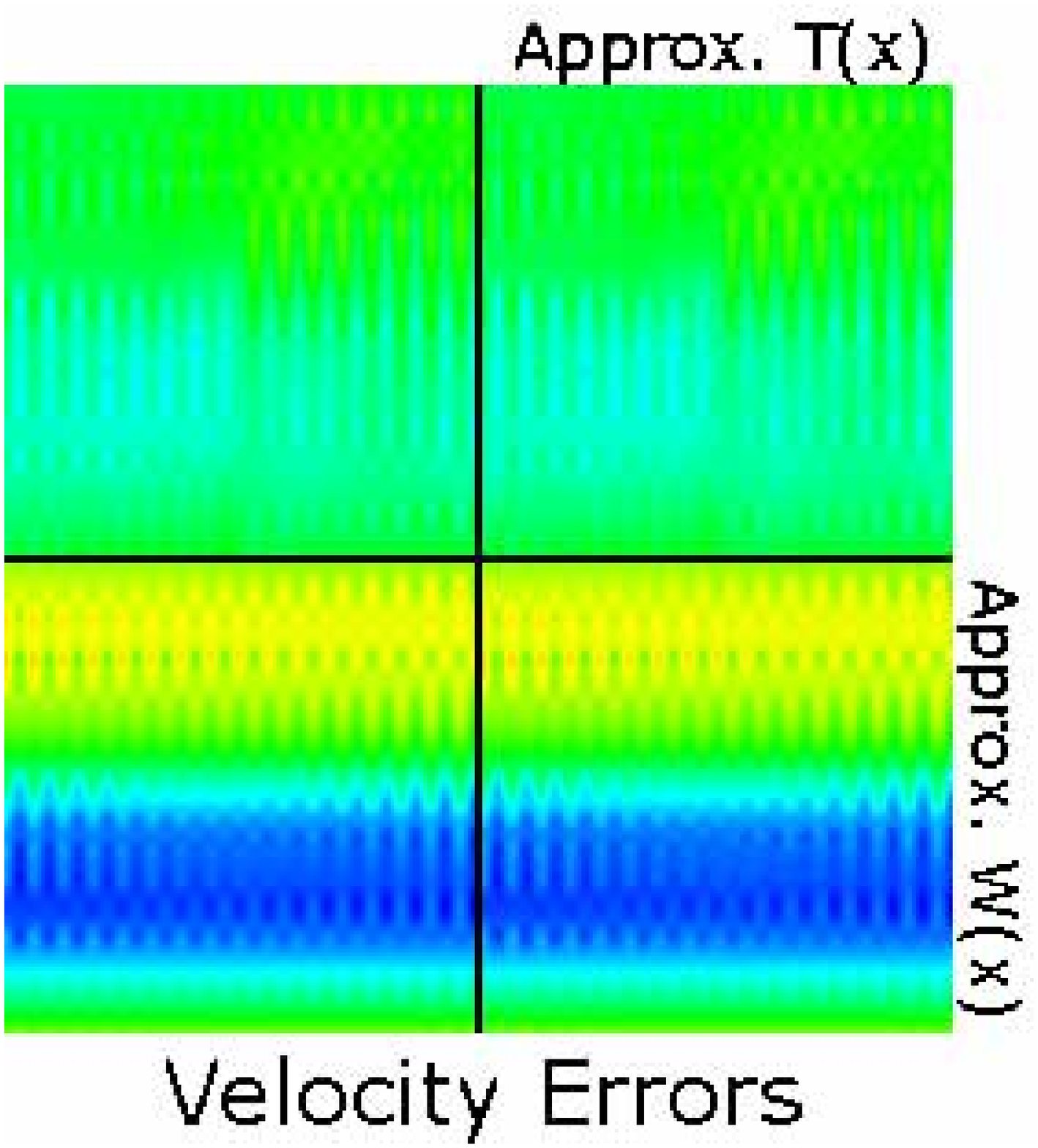}
  \caption{Errors in the mesh-refined density and velocity fields obtained
  by subtracting the upper left quadrants of Figures \ref{fig:multi2}
  and \ref{fig:grafic1024}.  False colors are scaled to $\pm0.005$ standard
  deviations for the density errors and $\pm0.2$ standard deviations for
  the velocity errors.  Each map is a mosaic of 4 panels showing the errors
  resulting from the two major approximations used in the multiscale
  computation.  The right columns, labelled ``Approx. T(x),'' show the
  effect of the spherical transform method for computing the transfer
  functions.  The lower rows, labelled ``Approx. W(x),'' show the effect
  of the minimal $k$-space sampling method for computing the anti-aliasing
  filters.  The upper-left quadrants show the errors when exact (and
  computationally expensive) transfer and anti-aliasing filters are used
  while the lower-right quadrants show the errors in Figure
  \ref{fig:multi2}.  There are residual errors even with exact $T(\vec x\,)$
  and $W(\vec x\,)$  because of the spatial truncation of $W$.}
  \label{fig:errmosaic}
\end{figure}

To quantify these errors, in Figure \ref{fig:errmosaic} we show residuals
obtained by subtracting the exact maps from the multiscale maps for the
32 Mpc refinement subvolume.  A priori we expect three main sources of
error:
\begin{enumerate}
\item The use of the spherical method for fast computation of the
short-wavelength transfer functions;
\item The use of the minimal $k$-space sampling method for fast
computation of the long-wavelength anti-aliasing filters; and
\item Truncation of the anti-aliasing filter to perform the
convolution over a subvolume instead of the entire top grid.
\end{enumerate}
All three effects are visible in Figure \ref{fig:errmosaic}.  The rows
and columns that are not labeled use the exact filters but are still
subject to the third error, truncation of $W(\vec x\,)$.

Scaled to the standard deviation of the high-resolution density field,
the rms errors of density in the subvolume shown in Figure
\ref{fig:errmosaic} are 0.04\% (upper left), 0.09\% (upper right),
0.06\% (lower left), and 0.10\% (lower right).  Thus, the major source
of error for adaptive refinement of the density field is the use of a
spherical transfer function for the short-wavelength components.  The
magnitude of the error is insignificant for the accuracy of cosmological
simulations.  For the velocity field, on the other hand, the corresponding
rms errors are 3.2\% (top row) and 7.0\% (bottom row).  Clearly the
anti-aliasing filter step is causing problems for the long-wavelength
velocity field.

\subsection{Solving the Anti-aliasing Problems for the Velocity Field}
\label{sec:fixv}

The long-range coherence of the velocity (or gravity) field has been seen
to cause difficulties for the evaluation of the long-wavelength components
by anti-aliasing the coarse-grid sample.  This subsection presents a
solution.

Several attempts were made to reduce the anti-aliasing errors while
continuing to use a minimal $k$-space sampling algorithm.
None of the attempts succeeded until we split the long-wavelength velocity
field from the top grid into parts due separately to the mass inside and
outside the refinement subvolume.  The motivation for this was the idea
that the latter part (the tidal field within the subvolume caused by mass
outside it) might be smooth enough to require minimal interpolation
to the subgrid.  For convenience, tidal split was done by setting $\xi_0
(\vec m\,)=0$ outside or inside the subvolume instead of setting $\delta_0
(\vec m\,)=0$; the coherence length of the density field is so small that
very little difference is made either way.  Linearity of the velocity
field ensures that when we add together the two parts either way we get
the complete long-range velocity field.

\begin{figure}[t]
\includegraphics[scale=0.4]{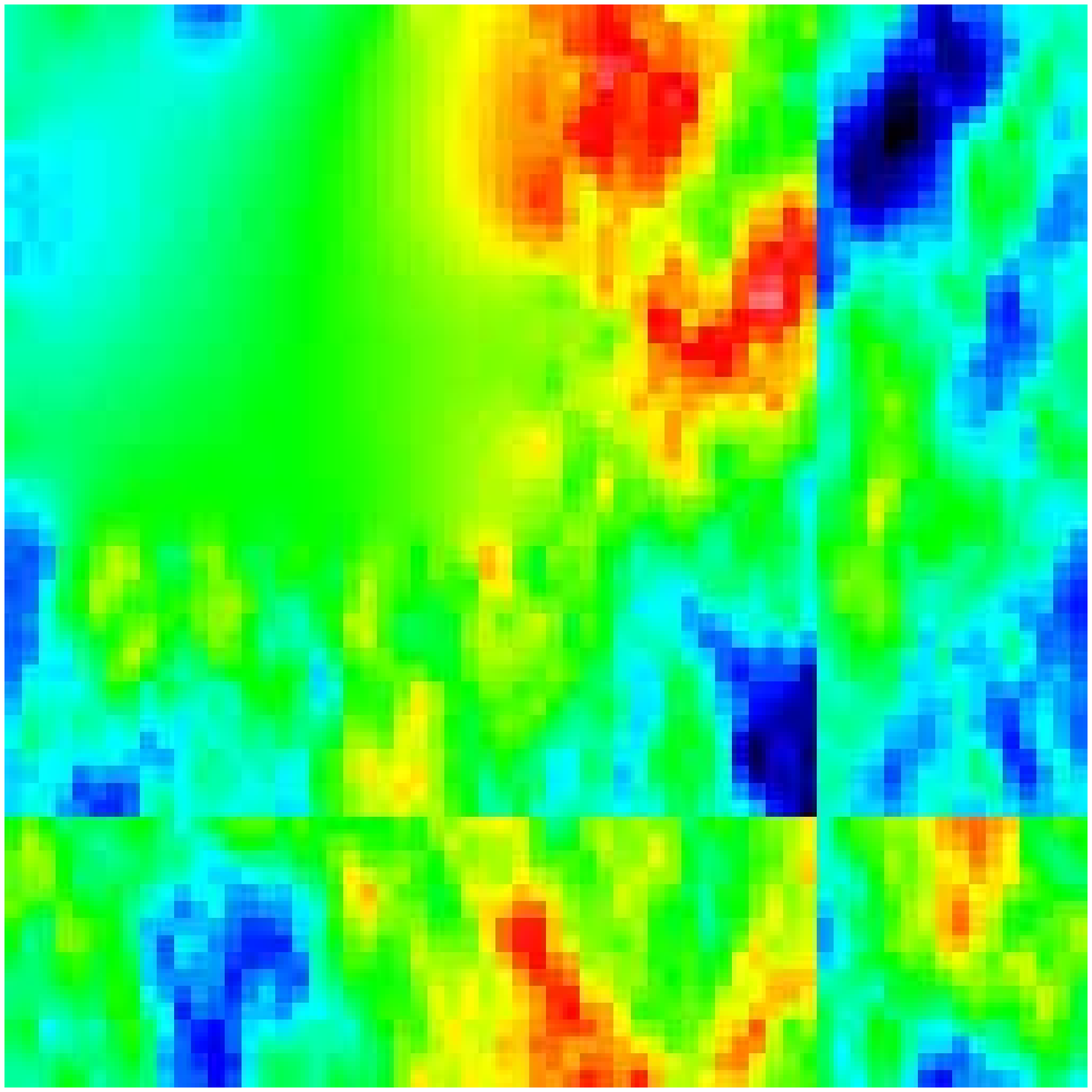}
\includegraphics[scale=0.4]{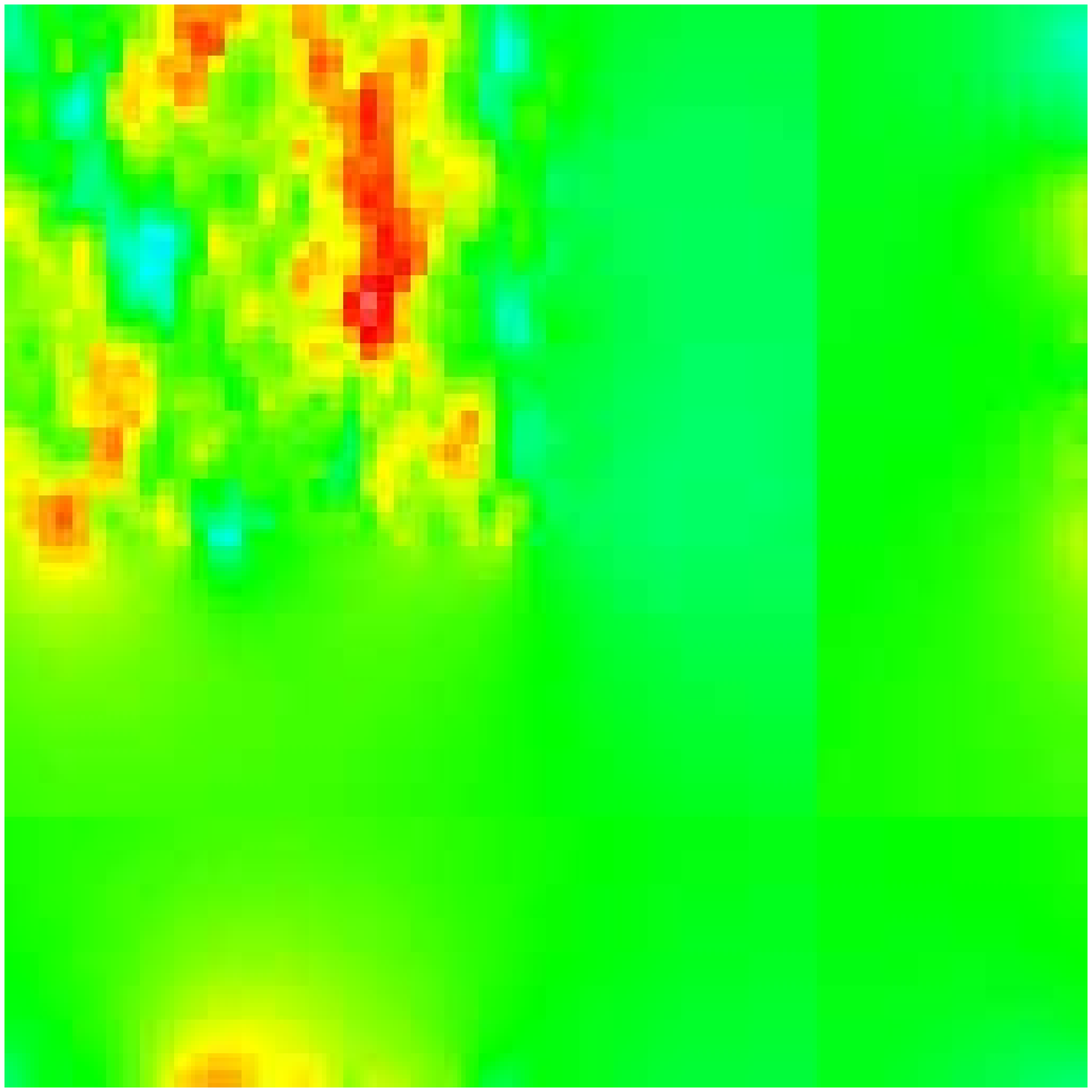}
  \caption{Split of the coarse-grid velocity field into parts due to
  fluctuations outside (left, ``outer part'') and inside (right,
  ``inner part'') of the refinement subvolume.  The two maps together
  add to give the velocity map of Figure \ref{fig:delvlong}.  This
  decomposition is the key to reducing the anti-aliasing velocity
  field errors, as described in the text.}
  \label{fig:tides}
\end{figure}

Figure \ref{fig:tides} shows the decomposition of the velocity field into
the ``outer'' and ``inner'' parts.  They were computed by zeroing the
white noise field in the appropriate regions and re-running {\tt GRAFIC}.
The same boundary conditions are used as in Figure \ref{fig:delvlong}.
The character of the two parts is strikingly different within the
refinement subvolume (the upper left quadrant).  The outer part is
smooth, as expected.  The inner part has a smaller coherence length
and it is well-localized over the upper left quadrant.  This spatial
localization and coherence suggest that the truncated minimal $k$-space
filter will be much more accurate for the inner part than for the
complete velocity field.  For the outer part, on the other hand, we
know that the discontinuities at the boundary of the buffer regions
will cause appreciable errors if convolved with the same filter.
The smoothness of the tidal field inside the subvolume suggests that
we use a much simpler and more localized filter.

\begin{figure}[t]
\includegraphics[scale=0.8]{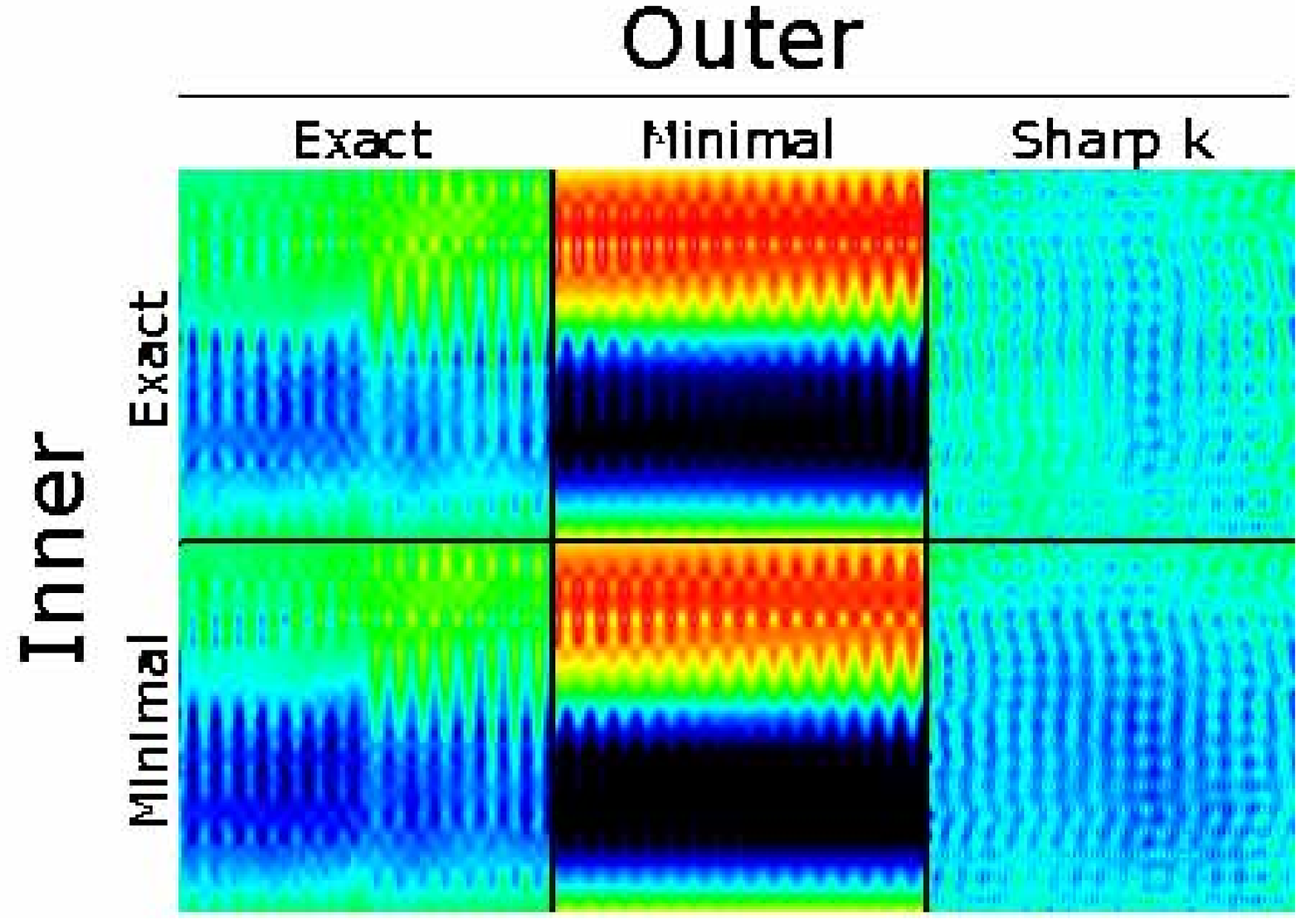}
  \caption{Velocity field errors under different approximations for
  evaluation of the inner and outer parts shown in Figure \ref{fig:tides}.
  The false colors are scaled to $\sigma/10$ where $\sigma$ is the
  standard deviation.  The upper left (exact/exact) and lower middle
  (minimal/minimal) maps are the same as the two rightmost maps in
  Figure \ref{fig:errmosaic}, where they were imaged with a color stretch
  only half as large ($\sigma/5)$.  RMS errors for each map (as a
  percentage of the rms one-dimensional velocity) are 3.2, 6.8, 2.7
  (top row, left to right) and 3.6, 7.0, 3.2 (bottom row).  The bottom
  right map gives the errors for the best fast method.  As in Figure
  \ref{fig:errmosaic}, there are errors even in the ``exact'' case
  because of the truncation of the anti-aliasing filter.}
  \label{fig:vxerr3}
\end{figure}

Several different filters were tried for the outer (tidal) part of the
velocity field.  The results for three are shown in Figure \ref{fig:vxerr3}.
The best simple filter was found to be sharp $k$-space filtering, which
sets $W(\vec k)=0$ everywhere except the fundamental Brillouin zone, where
$W(\vec k\,)=1$ (before the Hanning filter applied at the fine mesh scale).
This filter completely eliminates the aliasing error by eliminating the
replication of the fundamental Brillouin zone in $k$-space.  It is also
much more localized than the exact filter, so that spurious effects from
the buffer truncation in the left panel of Figure \ref{fig:tides} are not
convolved into the subvolume.  The price one pays is that it has the wrong
shape at small distances compared with the exact filter, leading to a new
source of errors in the rightmost columns of Figure \ref{fig:vxerr3}.
However, these errors are smaller than the error made with the exact filter
due to the buffer truncation (top left panel).

A comparison of the top and bottom rows of Figure \ref{fig:vxerr3} shows
that the filtering of the inner part of the velocity field is a minor
source of error.  It is the tidal field (the outer part) that requires
delicate handling.  Using a sharp $k$-space filter for the outer part
and minimal $k$-space filter for the inner part, our final errors are
3.2\% rms, the same as if we had used the computationally expensive
exact filter throughout.  These errors are probably small enough to be
unimportant in cosmological simulations.  They could be further reduced,
at the expense of an increase in computer time and memory, by increasing
the size of the buffer region for the top grid.

\section{MULTIPLE REFINEMENT LEVELS}
\label{sec:multiple}

The ability to refine an existing mesh opens the possibility
of recursive refinement to multiple levels, offering a kind
of telescopic zoom into cosmic structures.  Before this digital
zoom lens can work, however, there are some implementation issues
to face.  The issues addressed in \S \ref{sec:test} must be
considered anew in light of recursive refinement.

To see the issues arising in recursive refinement, consider a
three-level refinement with refinement factors $r_1$ and $r_2$.
By analogy with equations (\ref{xgrid}) and (\ref{xisamp}), we write
the grid coordinates and noise fields as
\begin{equation}
  \label{xgrid3}
  \vec x(\vec m,\vec n,\vec o\,)=\vec x_o+\left(L\over M\right)\left
    (\vec m+{1\over r_1}\vec n+{1\over r_1r_2}\vec o\,\right)\ .
\end{equation}
\begin{equation}
  \label{xisamp3}
  \xi(\vec m,\vec n,\vec o\,)=\xi_2(\vec m,\vec n,\vec o\,)+
    \xi_1(\vec m,\vec n\,)-\bar\xi_2(\vec m,\vec n\,)+\xi_0(\vec m\,)-
    \bar\xi_1(\vec m\,)
\end{equation}
where $\bar\xi_2$ is obtained by averaging $\bar\xi_2$ over $\vec o$.
At each level of the hierarchy there is a different grid (labeled by
$\vec m$, $\vec n$, and $\vec o$, respectively).  The variances
of the white noise samples are related by ${\rm Var}(\xi_2)=r_2^2
{\rm Var}(\xi_1)=r_1^2r_2^2{\rm Var}(\xi_0)$.

The main idea of recursive refinement is that, once we have refined
to level $n$ (where $n=0$ is the periodic top grid before any
refinement), the fields computed at that level serve as top-grid
fields to be refined to level $n+1$.  Equations (\ref{del12}),
(\ref{del12a}), and (\ref{tildel2}) showed how that refinement works
for $n=0$ by applying an anti-aliasing filter to the level-0 fields.
For $n=1$ we get
\begin{equation}
  \label{del3}
  \delta(\vec m,\vec n,\vec o\,)=\delta(\vec m,\vec n\,)\ast W+
    \left[\xi_2(\vec m,\vec n,\vec o\,)-\bar\xi_2(\vec m,\vec n\,)
    \right]*T\ .
\end{equation}
The procedure for refinement to an arbitrary level is now clear.
First we sample the fields at the preceding level and spread them
to the new fine grid.  Then we convolve with the appropriate
anti-aliasing filter.  Next we sample short-wavelength noise
on the new fine grid and subtract the coarse-cell means so that
so that the noise is zero at every higher level of the hierarchy.
This noise is then convolved with the transfer function and
added to the long wavelength field to give the high-resolution field.
This procedure is the same for all levels of the hierarchy.  However,
there are some issues to consider involving the transfer functions
and anti-aliasing filters.  We discuss these next.

\subsection{Short Wavelength Components}
\label{sec:multishort}

As was the case for two-level refinement, exact sampling requires that
we compute the upper-level sample without any filtering.  That is, we
should eliminate the Hanning filter from both the anti-aliasing filter
$W$ and the transfer function $T$ before computing all refinements
except the last one at the highest degree of refinement.  Otherwise
we would lose power present in the intermediate refinement levels.

Eliminating the Hanning filter is straightforward for the anti-aliasing
filters used in \S \ref{sec:implem}.  The minimal $k$-space and sharp
$k$-space filters are equally easy to compute with or without a Hanning
filter.  However, the transfer functions are an altogether different matter.
In \S \ref{sec:short} we used spherical transfer functions after concluding
in \S \ref{sec:minimal} that the coarse $k$-space sampling of the minimal
method would give significant errors for the short-wavelength fields.

Unfortunately, the unfiltered density transfer function is anisotropic,
as was shown in Figure \ref{fig:transd3}.  It also has a higher peak
value than the filtered transfer function in Figure \ref{fig:transf}.
Computing the exact transfer function is unacceptably costly, with
the operations count scaling as the sixth power of the total refinement
factor (i.e. the product of the individual refinement factors for each
level).  Thus, we are forced to reconsider the spherical and minimal
sampling methods for the transfer functions.

The unfiltered density is nonspherical because $k$-space is sampled in
a cube instead of a sphere.  Besides creating anisotropy, this sampling
increases the small-scale power.  As an alternative, we might use the
spherical method of equation (\ref{transkx3}) with a Hanning filter but
with a maximum spatial frequency (i.e. the cutoff for the Hanning filter)
larger than the Nyquist frequency $\pi/\Delta x$ for grid spacing $\Delta x$.
This is easily done by increasing the Nyquist frequency by a factor $f>1$
to include the high-frequency waves in the Brillouin zone corners with
$\vert\vec k\vert>\pi/\Delta x$.  For $f=1.838$ and the power spectrum
parameters used before, this method reproduces the correct peak value of
the density transfer function.  However, the spherical method cannot
reproduce the anisotropy evident in Figure \ref{fig:transd3}, which is
important on small scales.  (For multilevel refinement, exact sampling
requires that we use the anisotropic transfer functions at all but the
finest refinement.  The refinement process itself magnifies cubical pixels.
This anisotropy is cancelled by summing over the contributions from
different levels of the hierarchy only if we use anisotropic filters.)
On the other hand, the velocity transfer function is an integral over
the density transfer function and therefore is much smoother.  Errors at
small $r$ due to the neglect of anisotropy are much less important for the
velocity field.  Thus, we might approximate the correct, anisotropic
transfer function for the radial velocity by the spherical one.

\begin{figure}[t]
  \begin{center}
    \includegraphics[scale=0.6]{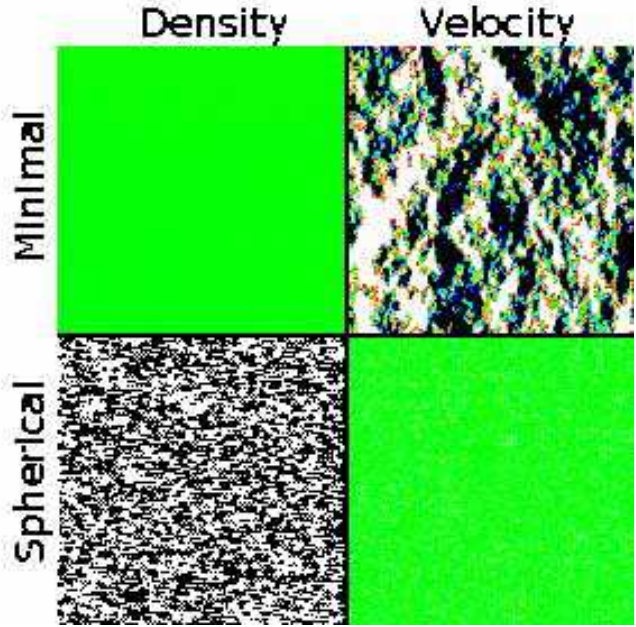}
  \end{center}
  \caption{Errors of the short-wavelength density (left) and velocity
  (right) fields computed without a Hanning filter, using spherical and
  minimal $k$-space filters as described in the text.  The density errors
  have been scaled to $\sigma/200$ and the velocity errors to $\sigma/10$.
  The minimal method is accurate at small spatial scales (density) while
  the spherical method is accurate at large scales (velocity).}
  \label{fig:dvnohan}
\end{figure}

Based on these considerations, we reconsider both the spherical and
minimal $k$-space sampling methods for approximating the unfiltered
transfer functions.  Figure \ref{fig:dvnohan} shows the residuals
from the exact, anisotropic transfer functions.  Comparison with Figure
\ref{fig:errmosaic} shows that removal of the Hanning filter adds no
significant errors provided that the minimal method is used for the
density and the spherical method is used for the velocity.  The
spherical method works poorly for the density field because it
neglects the small-scale anisotropy.  The minimal method works poorly
for the velocity field because it assumes periodicity on a scale
twice the subgrid extent.

The minimal sampling method works better than expected for the density
transfer function.  The reason for its success is that for CDM-like
power spectra the transfer function is dominated by large spatial
frequencies for which the coarse sampling of Fourier space introduces
little error in the discrete Fourier transform.  For the velocity
transfer function, long-wavelength contributions dominate and the
small-scale errors of the spherical method cause little harm.

\subsection{Long Wavelength Components}
\label{sec:multilong}

The next issue to consider is the treatment of tidal fields from
coarser levels of the grid hierarchy during the anti-aliasing step.
As we found in \S \ref{sec:fixv}, contributions from fluctuations
inside the subvolume can be filtered using the minimal $k$-space
sampling method, but contributions from tides generated outside the
subvolume must be convolved with a sharp $k$-space filter.  This
requires a clear separation of ``inner'' and ``outer.''  Care is
needed in the case of a multilevel hierarchy.

Consider, for example, refinement of the 256 Mpc top grid shown in
Figure \ref{fig:deltop} in two stages to produce the four-fold
refinement of the 32 Mpc level-2 subvolume in Figure \ref{fig:multi2}.
The level-1 subvolume may have any size between 64 and 128 Mpc.
(Each refinement must be over a subvolume no more than half the
size of the upper-level volume in order to accomodate the buffer
region used in the anti-aliasing step.)  When computing the
long-wavelength velocity contributions for the level-2 grid, the
level-1 fields must be computed with a tidal volume of 32 Mpc and
not the size of the level-1 subvolume.  Moreover, the same is true of
the level-0 fields.  Correct treatment of the tidal fields requires
that {\it all} upper levels be sampled with $\xi=0$ inside (or outside)
the final high-resolution subvolume.

This requirement implies that a chain of refinements must be performed
for every level of the hierarchy.  Computing a level-1 refinement
requires only one application of convolution plus small-scale noise.
Computing a level-2 refinement requires two applications: one to get
the level-1 samples with the correct tidal volume and a second to get
the level-2 results.  Thus, computing all three levels (0, 1, and 2)
requires three runs of the periodic grid routine in a box of size
256 Mpc (with no tides, with tides for the level-1 subvolume, and
with tides for the level-2 subvolume) plus three runs of the refinement
algorithm.

\begin{figure}[t]
  \begin{center}
    \includegraphics[scale=0.6]{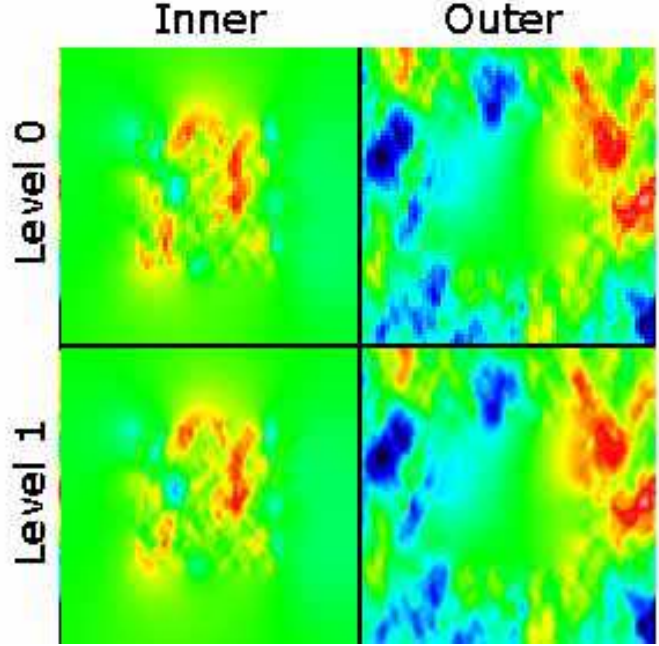}
  \end{center}
  \caption{Tidal fields in a volume 64 Mpc across from level 0 (top) and
    after refinement to level 1 (bottom).  This figure shows the successive
    refinement of the tidal fields needed for computation of the level-2
    velocity field.  The anti-aliasing has been performed without a
    Hanning filter in order to preserve exact sampling of the power spectrum.}
  \label{fig:tide2}
\end{figure}

The process of successive refinement is illustrated in Figure \ref{fig:tide2}
for the computation of the level-2 velocity field.  The top row is the same
as Figure \ref{fig:tides} except that the buffer has been unwrapped to
surround the volume.  However, instead of being prepared for a $r=4$
refinement, these top-level fields are prepared here for a $r=2$ refinement.
They are convolved with the appropriate anti-aliasing filters and short
wavelength noise is added to give level 1, shown in the lower row.
The level-1 tidal fields are resolved better and do not suffer from
aliasing at the resolution shown (0.5 Mpc grid spacing).  These fields
provide the input to a final $r=2$ refinement to produce the level-2 fields.

Comparing the resulting level-2 fields with Figure \ref{fig:grafic1024},
we find that the magnitude of the errors depends on the size of the level-1
grid.  For the case shown in Figure \ref{fig:tide2}, with a 64 Mpc grid,
the rms density and velocity errors are 0.02\% and 3.9\%, respectively.
When the level-1 grid size is increased to its maximum value of 128 Mpc,
these errors drop to 0.0094\% and 3.3\%, respectively.  These compare with
the errors for a single $r=4$ refinement, 0.10\% and 3.2\%, respectively.

The density errors have decreased with two $r=2$ refinements compared
with one $r=4$ refinement mainly because the minimal $k$-space sampling of
the anti-aliasing filter is coarser (hence less accurate) for $r=4$.  For
the velocity field the errors are dominated not by the coarse sampling
errors in $W(\vec x\,)$ but rather in the errors due to its truncation.
In other words, it is the discontinuities at the edge of the buffer region
(shown in Figure \ref{fig:tides}) that cause problems.  However, when
the top grid is refined by $r=2$ in a subvolume of half its size, the
doubling used for the convolution has a fortunate side effect: the buffer
region fills out the volume so that the entire top grid is included.
In this special case, which applies to our 128 Mpc level-1 grid, there are
no errors from periodic boundary conditions and the minimal $k$-space filter
is exact.  The errors arise almost exclusively from the second level of
refinement.  Thus, the two-level refinement in this case has the same
velocity field errors as a single $r=4$ refinement.

\section{MORE TRICKS WITH CONVOLUTION OF WHITE NOISE}
\label{sec:tricks}

The convolution method presented in this paper lends itself to a
variety of tricks that can be done with sampling of Gaussian
random noise.  These need not always involve adaptive mesh refinement
and convolution with isolated boundary conditions.  For example, in
Figure \ref{fig:grafic1024} we achieved multiscale initial conditions
using a single high-resolution grid but with the white noise sampled
more finely within a subvolume.  This procedure has the advantage of
allowing multiscale fields to be computed free from aliasing errors.
Although it is limited by computer memory constraints, this method is
the preferred choice for producing multiscale fields when computer memory
is not a limitation.

Our white noise sampling and convolution methods offer another way to
change the dynamic range of a simulation while retaining the sampling of
a fixed set of cosmic structures.  Instead of refining to small scales,
one may change the large scale structure in the simulation by expanding or
shrinking the top grid size.  This offers a simple and useful way, for
example, to add or subtract long waves in order to examine their effect
on small scale structure.  This brief section presents the method for
expanding or shrinking a simulation.

One way to implement this idea, which we will not explore further,
is to take an existing small-scale simulation to provide the
high-resolution field $\delta_1(\vec m,\vec n\,)$ as in equation
(\ref{del12}).  The small volume, originally with periodic boundary
conditions, is then embedded with isolated boundary conditions in a new
top grid field $\delta_0(\vec m\,)$.  The white noise sample used to
generate the existing small-scale simulation is taken to be $\xi_1
(\vec m,\vec n\,)$ and a new sample is created for the top grid.  This
procedure is the same as that described in \S \ref{sec:implem} except
that there the top grid sample was given and the subgrid sample was
added.  Here it is the other way around.  The implementation proceeds
as in \S \ref{sec:implem}.  It is straightforward and need not be
elaborated.

An alternative method is to change the size of an existing grid while
retaining a fixed grid spacing without refinement.  This method is easy
to implement because no aliasing occurs if the grid is not refined.
Moreover, periodic boundary conditions are used for all convolutions.
We simply change the scale of periodicity.  This can be achieved using
a modified version of the {\tt GRAFIC} code \citep{b95} called {\tt GRAFIC1}
which is being distributed along with the mesh-refinement version
{\tt GRAFIC2}.

The procedure is as follows.  First, identify a volume (perhaps a
subvolume of an existing simulation) whose size is to be changed.
{\tt GRAFIC1} should be run so as to output the white noise field
$\xi(\vec m\,)$ used in constructing the initial conditions.  Note that
the spatial mean noise level $\bar\xi$ vanishes so that the mean density
matches that of the background cosmological model.  This is a consequence of
periodic boundary conditions.

The white noise field is now expanded with the addition of new white
noise if one wishes to expand the box.  If one wishes to shrink the box
instead, then some of the noise field is excised.  The amplitude of the
white noise must be changed according to equation (\ref{wiener3}).
For example, if the grid size is doubled in each dimension, the existing
sample must be multiplied by $2^{3/2}$ and the added noise must have
the same variance.  These manipulations are easy to perform in real space.
The absence of any correlations for the white noise makes the treatment
of boundary conditions very simple.  Note that $\bar\xi$ must vanish on
the final grid because of periodic boundary conditions.  However, if the
volume has been expanded by a factor $f>1$, the mean value within the
original volume can be changed by adding a constant, e.g. a normal
deviate with variance $f$.

Finally, the new white noise field is now given as input to a second run of
{\tt GRAFIC1}, which calculates the density and velocity fields using
exact transfer functions.

\begin{figure}[t]
\includegraphics[scale=0.4]{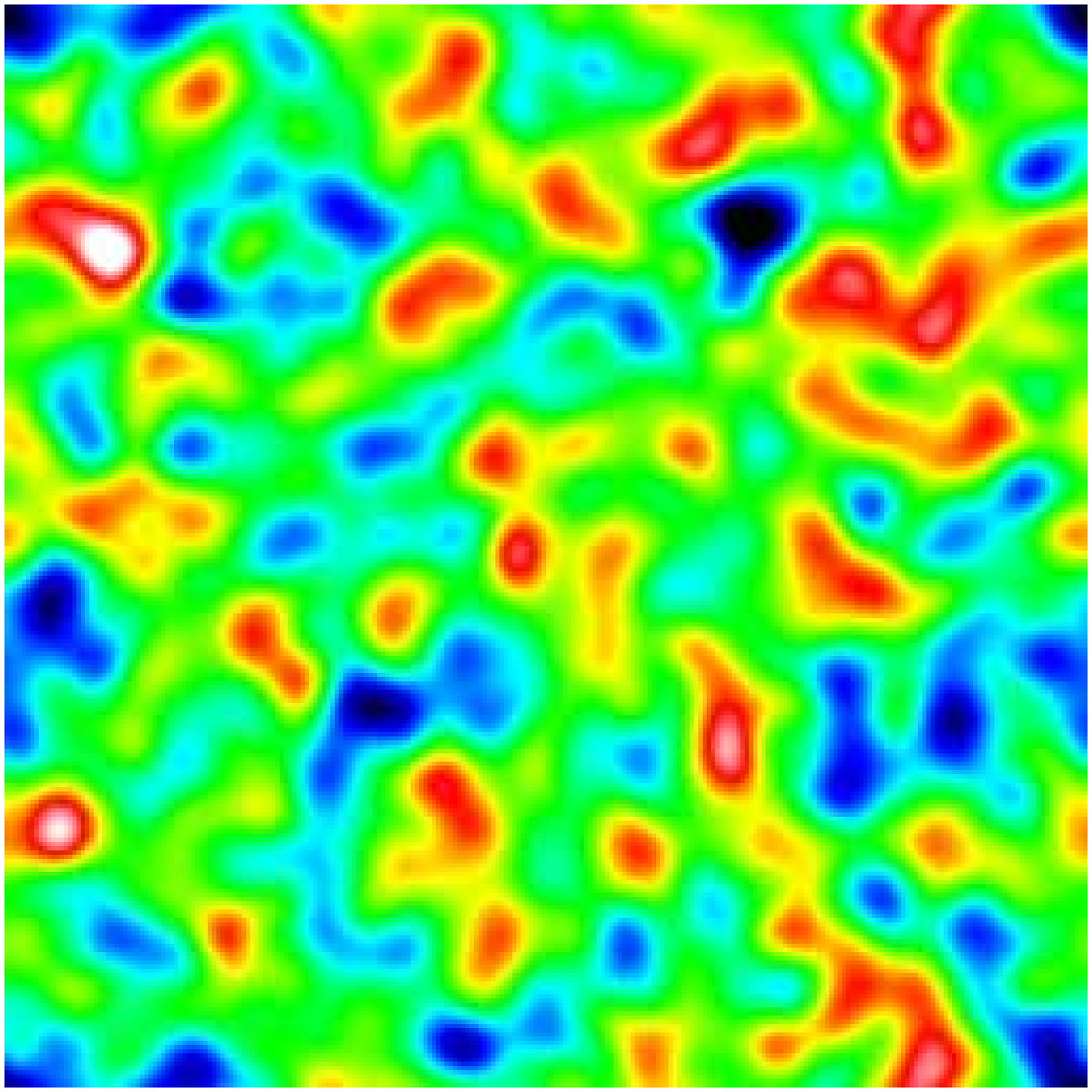}
\includegraphics[scale=0.4]{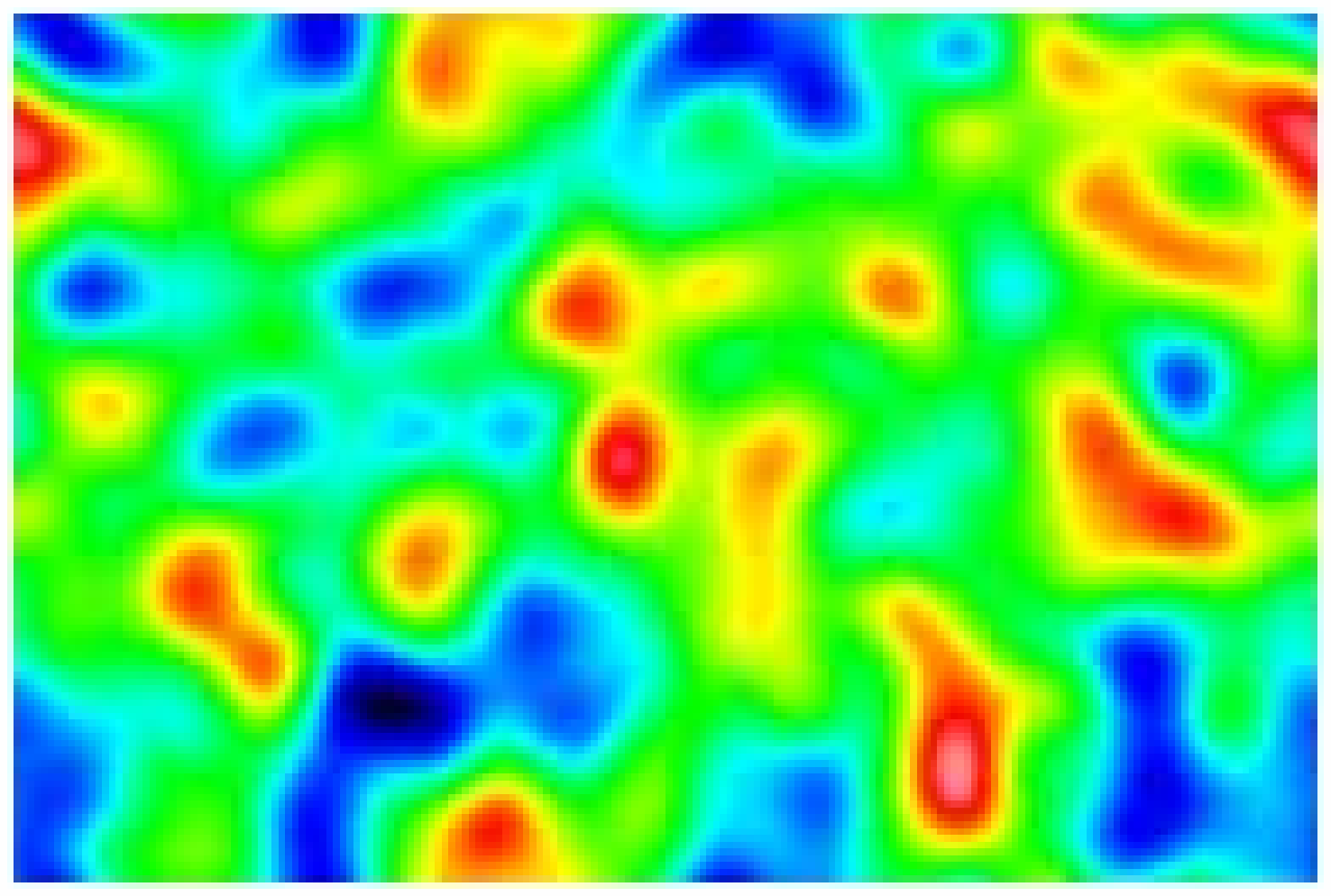}
  \caption{Slices of the density field from a realization of the Hot
  Dark Matter model.  Left: a cube of size 512 Mpc.  Right: a
  parallelpiped of dimensions $384\times256\times400$ Mpc has been
  extracted to create a new volume with periodic boundary conditions
  and with the same structures as the sample on the left.}
  \label{fig:shrinktop}
\end{figure}

This procedure is illustrated in Figure \ref{fig:shrinktop}.  One sees that
the structures in the left original volume are reproduced in the new sample
but that they are modified at the edges by the requirement of periodic
boundary conditions.  This affects the structure to within a distance of
a few coherence lengths.  The Hot Dark Matter model (with $\Omega_m=1$ and
$h=0.5$) was chosen for this test so that the coherence length would be
interestingly large.  One also sees that the initial conditions codes do
not require the volume to be a cube nor the axis lengths to be powers of 2.
{\tt GRAFIC1} and {\tt GRAFIC2} allow for arbitrary parallelpipeds as long
as there is at least one factor of two in each axis length.

\section{CONCLUSIONS AND CODE DISTRIBUTION}
\label{sec:end}

We have presented an algorithm for adaptive mesh refinement of Gaussian
random fields.  The algorithm provides appropriate initial conditions for
multiscale cosmological simulations.  Aside from small numerical errors,
the density and velocity fields at each refinement level are exact samples
of Gaussian random fields with the correct correlation functions including all
contributions from tides generated at lower-resolution refinement levels.
An arbitrary number of refinement levels is allowed in principle, enabling
cosmological simulations to be performed which have the correct sampling
of fluctuations over arbitrarily large dynamic ranges of length and mass.

Two convolutions are performed per refinement level for each field component.
These convolutions are performed using FFTs with the grid doubled in each
dimension.  Thus, the computer memory and time requirements for adaptive
mesh refinement are significantly greater than for sampling of Gaussian
random fields with a single grid.  One advantage of the refinement
algorithm is that the dynamic range in mass is not limited by the size
of the largest FFT that can fit into memory.  Also, it automatically
provides the correct initial conditions for multiscale simulations such
as that of \cite{abn}.

Adaptive mesh refinement of Gaussian random fields is more complicated than
refinement of, for instance, the fluid variables in a hydrodynamics
solver.  The reason for this is that Gaussian random fields have long-range
correlations.  Correct refinement within a subvolume cannot be done
independently of the lower resolution fields outside that subvolume.
When the resolution is increased by decreasing the pixel size, a given
sample suffers from aliasing.  Correct sampling requires convolution
by an anti-aliasing filter.  Short-wavelength contributions are then
provided by convolution of white noise with the appropriate transfer function.

Due mainly to imperfect anti-aliasing, numerical errors prevent one from
achieving perfect sampling of multiscale initial conditions.  However,
with careful analysis of the source of errors --- primarily from tides
generated outside the subvolume --- we have reduced these errors to an
acceptable level.  In testing with a realistic cosmological model, the
rms errors for a four-fold refinement were 0.1\% or smaller for the density
and 3\% for the velocity.  We showed that the most accurate results
are achieved by refinement factors of two, with each successive subvolume
occupying one-eighth the volume (half the linear extent) of the parent
mesh.  For a single refinement level, the anti-aliasing errors vanish
in this case.

Further testing is advised before the code is run to more than 4 refinement
levels or a total refinement greater than 16.  Also, some of the same
numerical issues (e.g. refinement of tidal fields) identified here may
arise in the gravity solvers used by nonlinear evolution codes.  Careful
testing of both the initial conditions and the nonlinear simulations codes
is advised before workers apply them to dynamic ranges in mass exceeding
$10^{11}$.  Unfortunately, it is very difficult to provide exact standards
for comparison with grid hierarchies of such large dynamic range.

The algorithm described in this paper has been implemented in FORTRAN-77 and
released in a publically available code package that can be downloaded from
{\tt http://arcturus.mit.edu/grafic/}.  The package has three main codes:
\begin{enumerate}
\item {\tt LINGERS} is an accurate linear general relativity solver
  that calculates transfer functions at a range of redshifts.
\item {\tt GRAFIC1} computes single-grid Gaussian random field samples
  with periodic boundary conditions.
\item {\tt GRAFIC2} refines Gaussian random fields starting with those
  produced by {\tt GRAFIC1}.  It may be run repeatedly to recursively
  refine Gaussian random fields to arbitrary refinement levels.
\end{enumerate}

{\tt LINGERS} is a modification of the {\tt linger\_syn} code from the
{\tt COSMICS} package \citep{b95,mabert95}.  It produces output at a range
of times enabling accurate interpolation to the starting redshift of the
nonlinear cosmological simulation.  {\tt CMBFAST} \citep{cmbfast} could
be used instead, although the treatments of normalization and units are
different for the two codes.

{\tt GRAFIC1} is a modification of the {\tt grafic} code from {\tt COSMICS}
that incorporates exact transfer functions for both CDM and baryons at
arbitrary redshift from {\tt LINGERS} and uses white noise sampled in real
space as the starting point for Gaussian random fields.  As demonstrated in
\S \ref{sec:tricks}, sampling of white noise enables one to change the
size of the computational volume, or to embed a given realization into a
larger volume with different resolution, simply by modifying the noise file.
{\tt GRAFIC1} and {\tt GRAFIC2} also have optional half-mesh cell offsets
for the CDM or baryon grids.

{\tt GRAFIC2} is the multiscale adaptive mesh refinement code.  It requires
substantial computing requirements for large grids, mainly because of the
need to double the extent of each dimension.  Thus, suppose that one has
a $256^3$ top grid and wishes to double the resolution in one-eighth of
the volume.  Computing the $256^3$ sample with {\tt GRAFIC1} requires
64 MB of memory.  Refining it with {\tt GRAFIC2} requires 1.02 GB of memory
and 3.5 GB of scratch disk.  The cpu time is also much larger, but is
still far less than the time required for the nonlinear evolution.
Fortunately, these computing resources are now available on desktop
machines.  Much larger grids are possible with parallel supercomputers.

I thank my colleagues in the Grand Challenge Cosmology Consortium ---
J. P. Ostriker, M. L. Norman, and L. Hernquist --- for encouragement in
this work.  Special thanks are given to L. Hernquist and the
Harvard-Smithsonian Center for Astrophysics for hosting my sabbatical
during this work.  Supercomputer time was provided by the National
Computational Science Alliance at the University of Illinois at
Urbana-Champaign.  Financial support was provided by NSF grants
ACI-9619019 and AST-9803137.

\end{document}